%% file: main_SIG19.tex
\documentclass[acmtog]{acmart}
\acmSubmissionID{207}

\newcommand{\revised}[1]{{\color{black} #1}}
\input{header}

\setcopyright{rightsretained}
\acmJournal{TOG}
\acmYear{2020}
\acmVolume{39}
\acmNumber{6}
\acmArticle{264}
\acmMonth{12} 
\acmDOI{10.1145/3414685.3417800}

\newcommand{\zo}{{\textsc{ZoomOut}}}

\begin{document}
\title{MapTree: Recovering Multiple Solutions in the Space of Maps}  

\author{Jing Ren}
\affiliation{\institution{KAUST}}
\email{jing.ren@kaust.edu.sa}

\author{Simone Melzi}
\affiliation{
    \institution{LIX, \'Ecole Polytechnique}}
\email{melzismn@gmail.com}

\author{Maks Ovsjanikov}
\affiliation{
  \institution{LIX, \'Ecole Polytechnique}}
\email{maks@lix.polytechnique.fr}

\author{Peter Wonka}
\affiliation{
  \institution{KAUST}}
\email{pwonka@gmail.com}
\renewcommand{\shortauthors}{Ren, Melzi, Ovsjanikov, and Wonka}	 
\begin{abstract}
In this paper we propose an approach for computing multiple high-quality near-isometric dense correspondences between a pair of 3D shapes. Our method is fully automatic and does not rely on user-provided landmarks or descriptors. This allows us to analyze the full space of maps and extract multiple diverse and accurate solutions, rather than optimizing for a single optimal correspondence as done in most previous approaches. To achieve this, we propose a compact tree structure based on the spectral map representation for encoding and enumerating possible rough initializations, and a novel efficient approach for refining them to dense pointwise maps. \revised{This leads to a new method}
capable of both producing  multiple high-quality correspondences across shapes and revealing the symmetry structure of a shape without a priori information. In addition, we demonstrate through extensive experiments that our method is robust and results in more accurate correspondences than state-of-the-art for shape matching and symmetry detection.
\end{abstract}

\begin{teaserfigure}
  \centering
  \begin{overpic}[trim=0.3cm 0.1cm 1.7cm 3.5cm,clip,width=1\linewidth,grid=false]{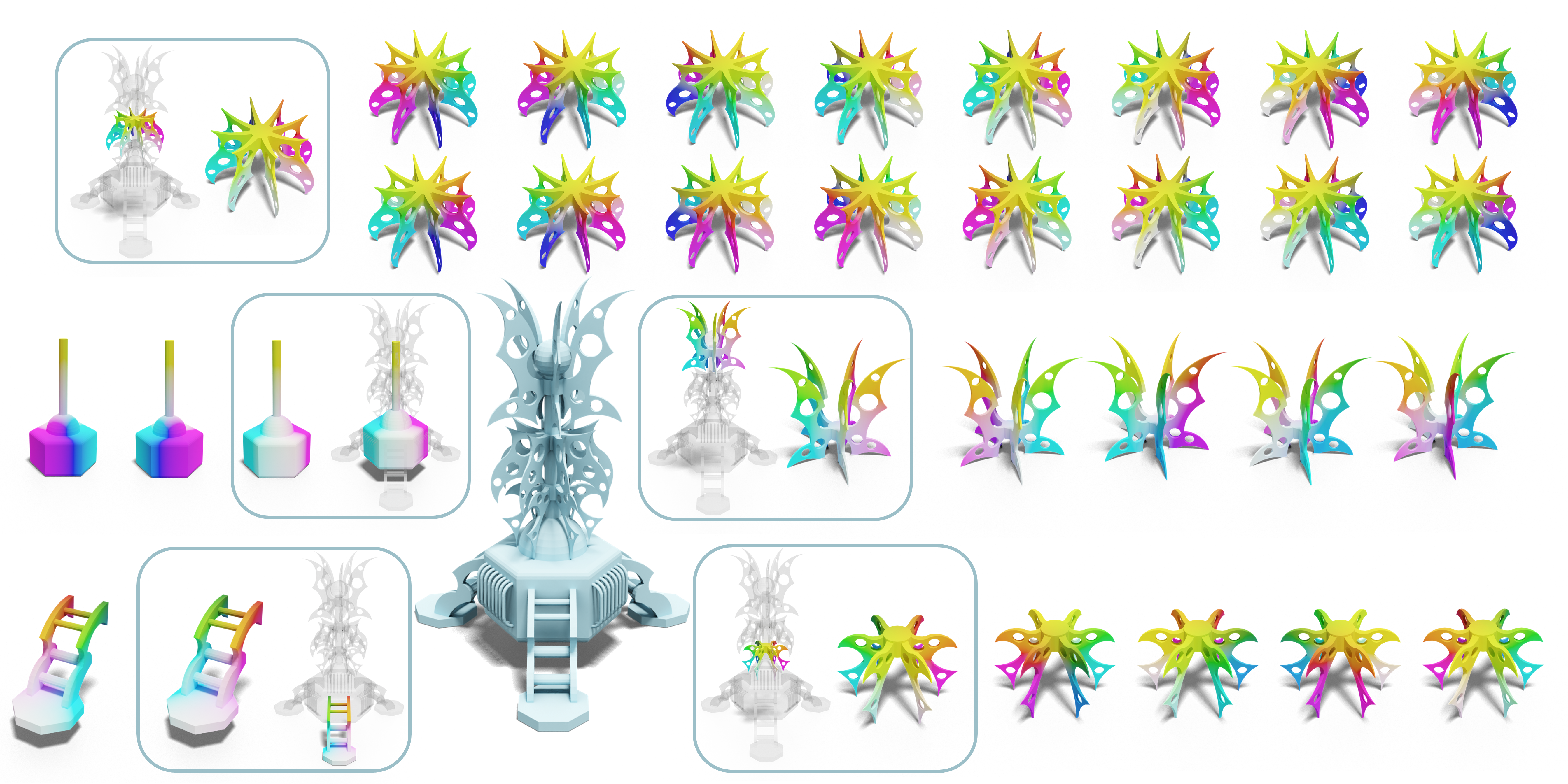}
 \end{overpic}
 \vspace{-20pt}
  \caption{For a given model (colored blue in the center), we first \revised{automatically} isolate its disconnected components (shown in boxes). For each component, we show its location on the original shape, and a zoomed-in version where we color the vertices using their 3D coordinates. On the left/right of the component, we show a \emph{subset} of the self-symmetric maps of this component, obtained fully automatically using our algorithm. Correspondences are shown via color transfer.}
 \label{fig:teaser}
\end{teaserfigure}


\maketitle

\section{Introduction}
Computing dense correspondences across non-rigid 3D shapes is a classical problem in computer vision, computer graphics, and related fields.
Dense correspondences between a pair of shapes are typically encoded by a map that assigns points on one shape to points on the other shape. There exist many optimization methods that define a map energy and try to find a desirable map by optimizing it~\cite{bronstein2006,huang2008,ovsjanikov2010,vestner2017product,ezuz2018reversible}.  Unfortunately, the space of maps is very complex.

\begin{figure}[!t]
    \centering
     \begin{overpic}[trim=0cm -1cm 0cm 0cm,clip,width=0.57\linewidth,grid=false]{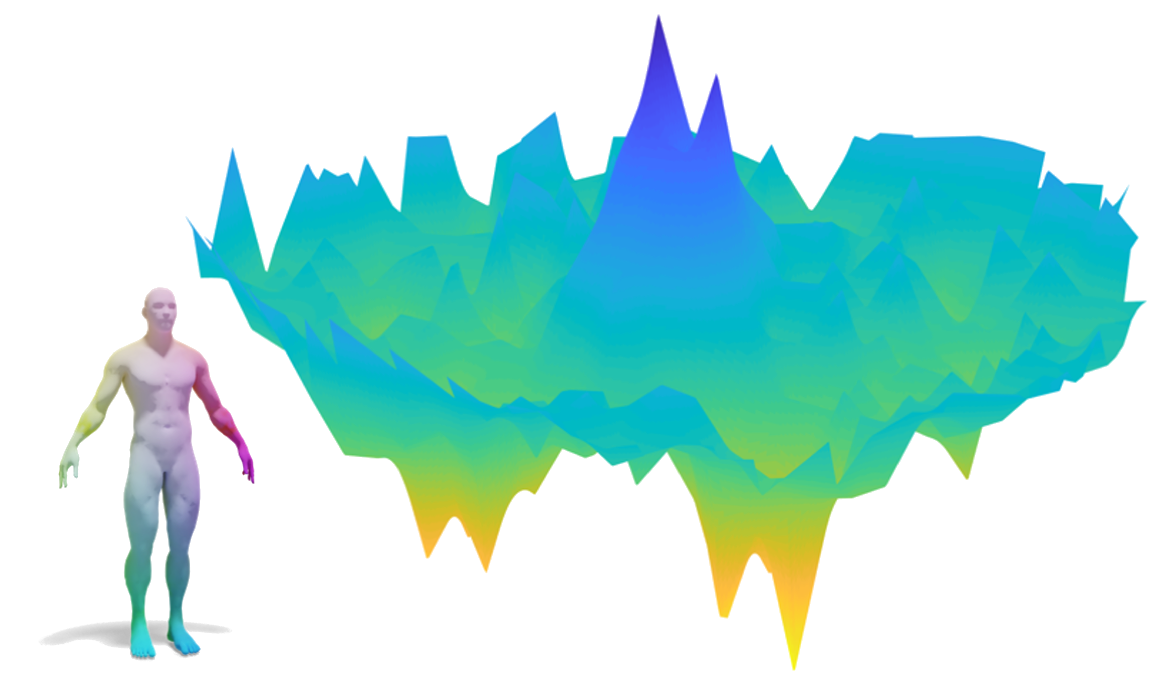}
     \put(18,5){\small (Source)}
     \end{overpic}\hspace{-2pt}
      \begin{overpic}[trim=0cm 2cm 0cm 2cm,clip,width=0.42\linewidth,grid=false]{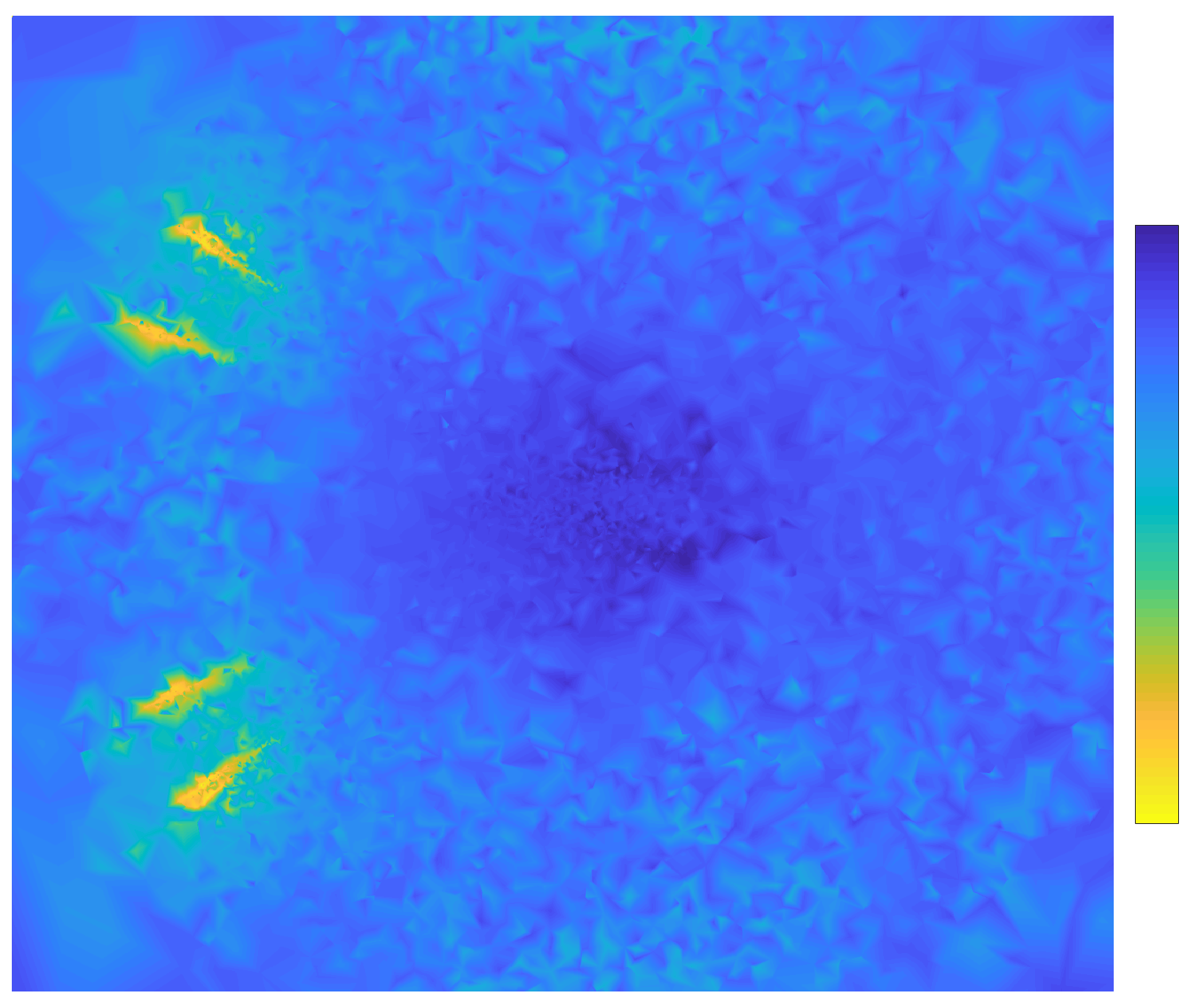}
     \put(10,82){\small Heatmap of the GeoDist}
     \put(-100,82){\small Landscape of the GeoDist}
     \put(95,3){\small Low}
     \put(95,65){\small High}
     \end{overpic}\vspace{-3pt}
     \caption{We generate 10K self-maps on a human shape (shown bottom left). We then embed these maps into 2D using multi-dimensional scaling such that the geodesic distance between two point maps is preserved by the Euclidean distance in 2D. Therefore, we can visualize the geodesic distortion ("GeoDist") of each map. \emph{Left:} the 3D landscape of the GeoDist over 10K maps. \emph{Right:} the corresponding 2D heatmap. The maps with small (large) geodesic distortion are in color yellow (blue).}
     \label{fig:landscape}
\end{figure}
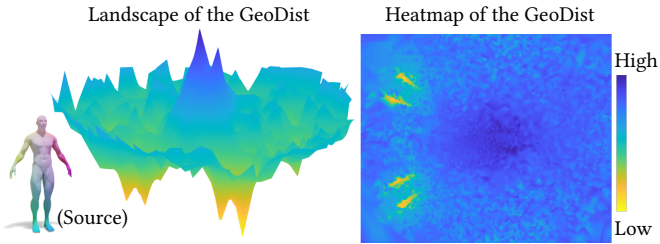

First, this space is exponential in size and can contain many local minima. For example, Fig.~\ref{fig:landscape} shows the commonly-used geodesic distortion (the average change in geodesic distances across all pairs of points) energy of 10K self-maps on a human shape. We can see that the energy function is non-convex with multiple local minima that correspond to different types of maps.
Some local minima correspond to semantically meaningful maps that exist due to the presence of symmetries, e.g. the left-to-right symmetry of a human shape. Other local minima correspond to other smooth high-quality maps, e.g. a twisted mapping where the lower part of a human is correctly mapped, but the upper part is flipped from left-to-right. In addition, there are local minima (typically with higher energy) that correspond to undesirable maps, e.g. a mostly smooth map, but with one large error such as the head being mapped to a hand.

Second, the optimization in the space of maps is difficult and current algorithms need to make many simplifications, e.g. via relaxations~\cite{fogel2013convex,solomon2016entropic,dym2017ds++}, early termination, etc.
As a result, algorithms may not actually converge to local minima, but terminate at points that may or may not be close to a local minimum. We call these points termination points (of a particular algorithm).

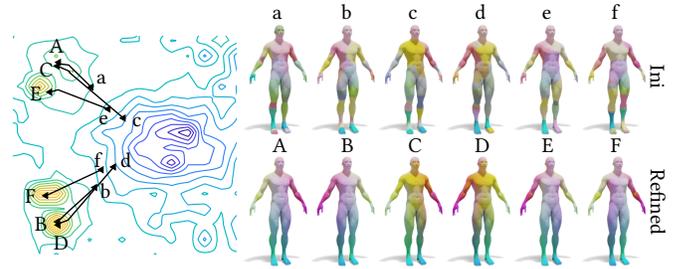
\begin{figure}[!t]
    \centering
    \vspace{-6pt}
     \begin{overpic}[trim=0cm -0.75cm 0cm 0cm,clip,width=0.35\linewidth,grid=false]{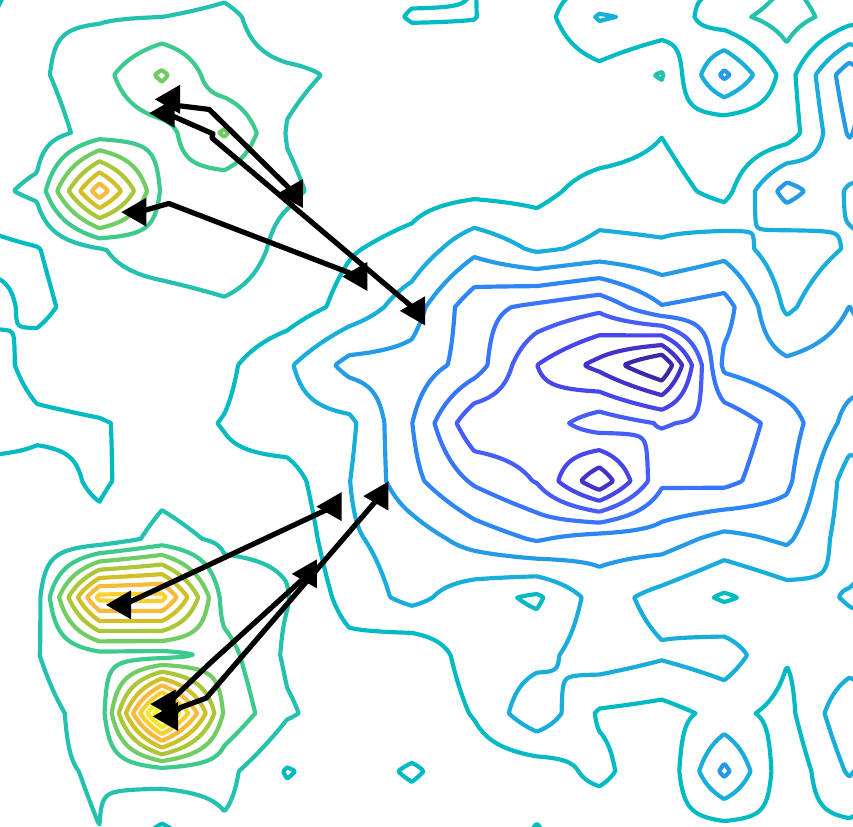}
     \put(15,93){\small A}  \put(35,80){\small a}
     \put(11,83){\small C}  \put(50,62){\small c}
     \put(7,73){\small E}   \put(36,63){\small e}
     \put(5,30){\small F}   \put(34,45){\small f}
     \put(9,18){\small B}   \put(36.5,32){\small b}
     \put(17,10){\small D}   \put(45,45){\small d}
     \end{overpic}
     \begin{overpic}[trim=1cm 0cm 1cm 0cm,clip,width=0.65\linewidth,grid=false]{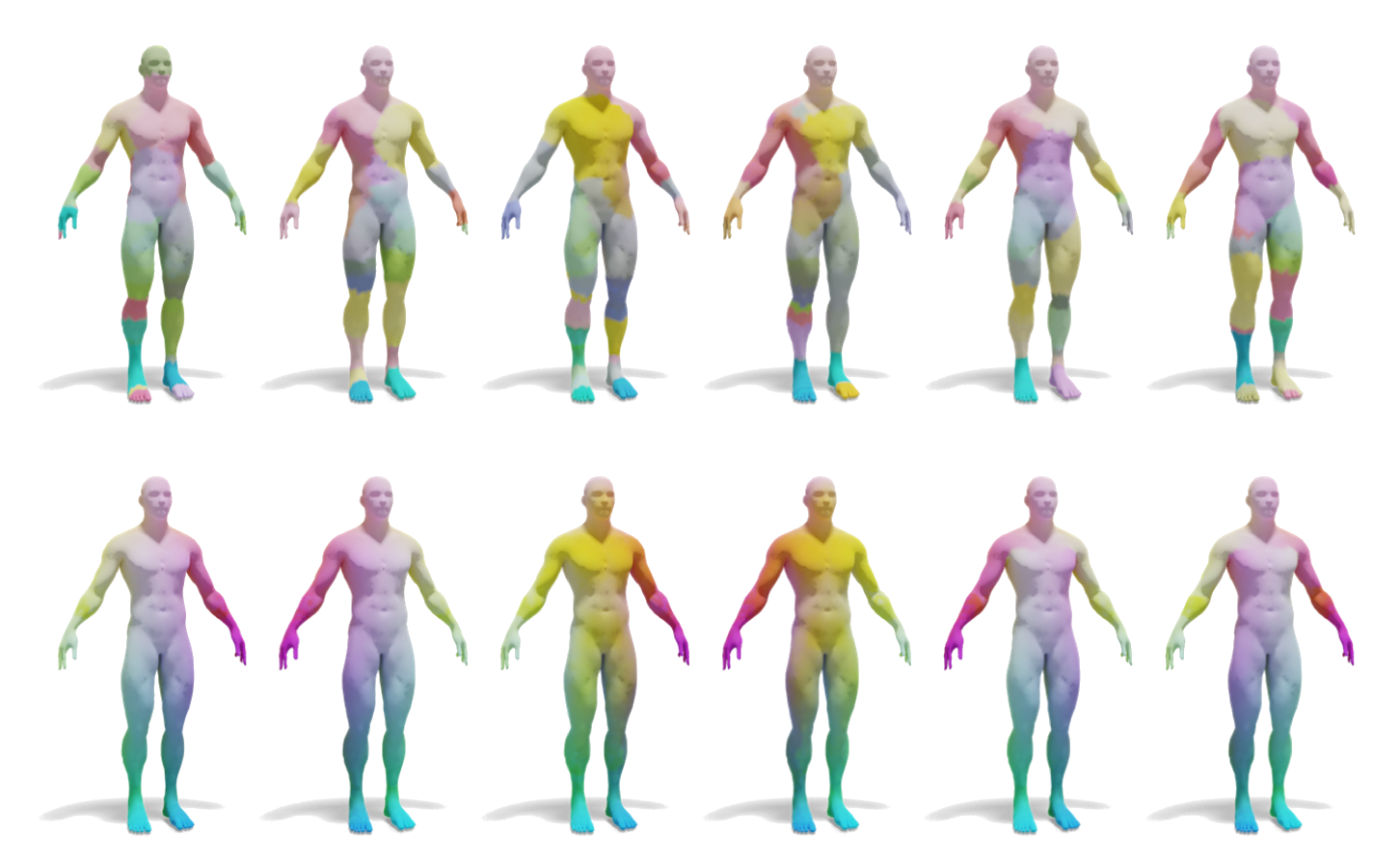}
     \put(99,50){\rotatebox{-90}{\small Ini}}
     \put(99,25){\rotatebox{-90}{\small Refined}}
     \put(8.5,61){\small a}     \put(8.5,29){\small A}
     \put(25,61){\small b}      \put(25,29){\small B}
     \put(41,61){\small c}      \put(41,29){\small C}
     \put(57,61){\small d}      \put(57,29){\small D}
     \put(73,61){\small e}      \put(73,29){\small E}
     \put(89.7,61){\small f}    \put(89.5,29){\small F}
     \end{overpic}\vspace{-9pt}
     \caption{\emph{Left:} the contour of the geodesic distortion energy as shown in Fig.~\ref{fig:landscape}. We pick 6 initial maps (a-f) and after applying ZoomOut we obtain the refined maps (A-F), the visualization of these maps are shown on the \emph{right}. The black lines show the intermediate steps/maps during the refinement/optimization. We can see that different initializations lead to different termination points (local minima).}
     \label{fig:contour}
\end{figure}

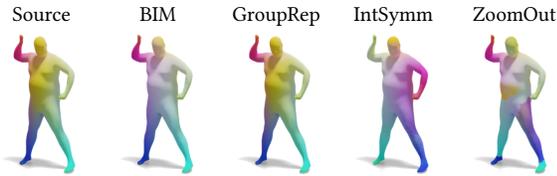
\begin{figure}
    \centering
     \begin{overpic}[trim=0cm 0cm 0cm 0cm,clip,width=0.9\linewidth,grid=false]{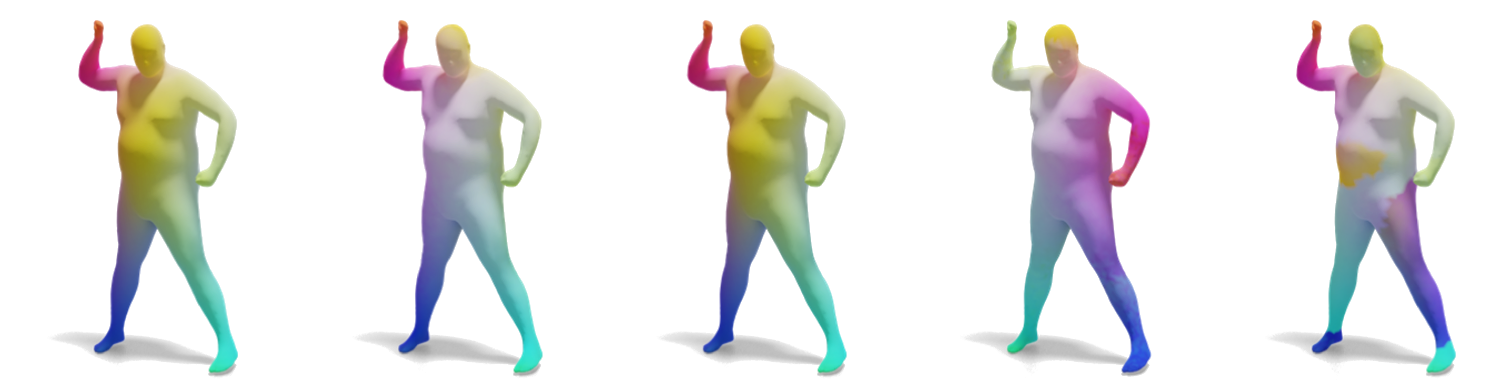}
    \put(5,27){\small Source}
    \put(27,27){\small BIM}
    \put(43,27){\small GroupRep}
    \put(64,27){\small IntSymm}
    \put(84.5,27){\small ZoomOut}
    \end{overpic}\vspace{-9pt}
    \caption{Self-symmetry detection. Here we show an example where different baselines lead to different local minima. Specifically, BIM~\cite{kim2011blended} obtains a back-to-front map, GroupRep~\cite{wang2017group} leads to the direct map, IntSymm~\cite{Nagar_2018_ECCV} leads to a back-to-front and left-to-right map, while ZoomOut~\cite{zoomout} that is initialized by OrientRev~\cite{ren2018continuous} obtains a map with mixed symmetry.}\vspace{-3pt}
    \label{fig:local_minima_method}
\end{figure}

Therefore, we can identify two types of errors in existing shape matching algorithms. 1) The algorithm terminates close to the correct local minimum, but it does not fully reach it because of the simplifications in the algorithm. This typically results in many small errors at each vertex, but only very few large errors.
2) The optimization algorithm ends up in a wrong part of the space, generally, because of the initialization. This latter problem is particularly prominent as many algorithms are based on pre-defined descriptors or landmarks ~\cite{vestner2017efficient,ovsjanikov2012functional,ren2018continuous}.
In Fig.~\ref{fig:contour} we visualize some example maps (A-F) that are results of a map optimization algorithm, ZoomOut~\cite{zoomout}, starting from different initializations.
In the example, four maps are recovered corresponding to symmetries: the direct map (A), the symmetric map (B), the back-to-front flipped map (C), and the doubly flipped map (D). In addition, there are two twisted maps. In map (E) the upper body is assigned the symmetric counterpart, while the the lower body arises from the direct map. In map (F) the upper body arises from the direct map and the lower body from the symmetric map.
In addition, when comparing different map optimization algorithms, one can observe that different methods can lead to different maps, see Fig.~\ref{fig:local_minima_method}.

We would like to advocate a departure from previous work, that has \revised{primarily} focused on recovering a single high-quality map, and suggest that recovering multiple (or all) high quality maps in the space of maps is an important goal.
Recovering multiple maps can provide information about all the intrinsic symmetries of the shapes and they also provide insights into the energy landscape of the optimization. For example, surprising low energy solutions of the geodesic distortion energy like the twisted maps can be recovered (not only accidentally) by existing methods. \revised{We therefore consider the problem of \emph{multi-solution shape matching} and derive a new solution to this problem.} We also note that multi-solution \emph{symmetry detection} is a special case of this problem (See Fig.~\ref{fig:teaser}).
However, our algorithm for recovering multiple maps is more than just a solution to a new problem statement. It can be used as building block to improve upon a long standing problem in traditional single-solution shape matching. For the vast majority of existing methods, the initialization is a considerable source of error and it can lead to a solution in the wrong part of the search space, e.g., a symmetric or twisted map is found when the goal is to compute the direct map.
Complementary to efforts in improving the initialization or the optimization, we demonstrate that a considerable improvement in single-solution shape matching can be achieved by first computing a set of all high-quality maps and then selecting among them. As we show below, this can lead to significant improvements over the state-of-the-art even in traditional single-solution shape matching and single-solution symmetry detection.


For designing an algorithm to recover multiple high-quality maps, we can identify two criteria that an ideal solution should satisfy: (1) the recovered maps should be sufficiently different from each other. (2) each dense map should be smooth and near isometric. This includes maps that correspond to symmetries and other globally smooth maps that minimize energies that promote near isometries, e.g., based on geodesic distortion.


To tackle this problem, we propose to exploit the \emph{spectral properties} of correspondences, encoded through the functional map representation \cite{ovsjanikov2012functional}. 
Key to our approach is a novel multi-resolution functional map exploration mechanism that completely avoids the need for descriptors or landmark correspondences and is capable of producing \emph{multiple} meaningful dense correspondences that might exist in particular due to symmetries on the shapes. Our method is based on three observations, which we discuss below in detail: 1) pointwise maps that have large differences differ in the low frequencies of the spectral basis; 2) maps that are globally similar mainly differ in the high frequencies; 3) Large functional maps can be recovered from a small functional map~\cite{zoomout}.   
Therefore, we explore the space of solutions through an efficient analysis of low frequency functional maps, which we then refine to obtain very accurate final correspondences.
Depending on the application the recovery of multiple maps may be complemented by an algorithm for map selection, e.g., in single-solution shape matching the direct map should be selected. 

To summarize, our main contributions are:
\vspace{-1.5mm}
\begin{enumerate}
    \item \revised{A new method} to solve the multi-solution shape matching problem that is able to output multiple high-quality dense correspondences.
    \item Significantly improved performance in near-isometric shape matching and symmetry detection compared to the state of the art.
    \item A compact tree structure for enumerating and organizing the space of maps between a pair of shapes, that allows exploration and map selection \emph{a posteriori}. 
\end{enumerate}


\section{Related Work}
\label{sec:related_work}
Below, we briefly review work that is most related to ours, while focusing on intrinsic symmetry detection, spectral methods and shape matching techniques.

\subsection{Intrinsic Symmetry detection}
In computer vision and computer graphics, intrinsic symmetries have been widely investigated as a tool to gain insight into the structure of shapes or scenes. We refer the interested readers to recent surveys~\cite{mitra13,Liu10}.

While early symmetry detection work primarily focused on \emph{extrinsic} or embedding-dependent symmetries \cite{mitra2006partial}, a significant number of techniques have also been proposed for \emph{intrinsic} symmetry detection, including \cite{ovsjanikov08,lipman2010symmetry,raviv10,xu2009partial,xu2012multi} among others. Importantly, most early intrinsic symmetry detection methods either only detect similar parts, and thus do not produce dense correspondences \cite{lipman2010symmetry,xu2009partial,xu2012multi}, suffer from robustness issues \cite{ovsjanikov08} or result in difficult non-convex optimization problems \cite{raviv10}, making them sensitive to initialization and strongly relying on descriptors and subsampling.



More closely related to ours are recent methods based on advances in shape correspondence, especially the functional maps framework, including \cite{Liu15,wang2017group,ren2018continuous,Nagar_2018_ECCV}. These methods also aim to produce dense correspondences but in most cases require additional information such as detection of extremities \cite{Liu15} or \emph{a priori} knowledge of symmetry orbits of a specified landmark of the shape \cite{wang2017group}. The method of Nagar et al.~\cite{Nagar_2018_ECCV} that we compare to below, is related to our approach in its exploration of near-diagonal functional map structure, but relies on very sensitive eigenfunction sign detection.

\subsection{Shape Matching}
Our framework is also related to the problem of shape matching, and thus to  methods that look for dense correspondences between non-rigid 3D shapes. For an in-depth review of this area we refer the readers to~\cite{biasotti2016recent,tam2013registration}.
Several approaches to shape matching directly solve for correspondences between points on the two surfaces by minimizing an explicit energy, e.g., ~\cite{bronstein2006,huang2008,ovsjanikov2010}.
The main limitation of these methods is that they often lead to complex combinatorial problems. An alternative is to first map the shapes to a canonical domain (e.g. a sphere), and then solve for the correspondence between these parametric representations~\cite{Lipman2009,Aigerman15,tutte}, or blend across multiple such maps \cite{kim2011blended}.

One successful strategy for shape matching is to relax the search space from permutation matrices to a space more amenable to continuous optimization, e.g., doubly stochastic matrices~\cite{fogel2013convex,solomon2016entropic,dym2017ds++}, but also other successful relaxations exist~\cite{leordeanu2005spectral,solomon2012,kezurer2015tight,maron2016point,Azencot19}. This, however, requires a possibly error-prone projection step to compute an integer solution, and makes it difficult to compute \emph{multiple} correspondences, which might exist due to symmetries.

\revised{Our goal is to automatically and simultaneously produce multiple high-quality maps present due to different symmetries. Though there are no existing methods to solve our problem directly, there is some pioneering work for shape matching where multiple maps are \emph{involved}. For example, multiple maps are generated and then blended into a single map in~\cite{kim2011blended}.
  \cite{sahilliouglu2018genetic} proposed a genetic method for shape matching, where a ``population'' of maps are maintained for ``evolution''. } However, these approaches do not aim to produce multiple different high-quality dense correspondences and thus they cannot be easily adapted to our setting. \cite{sung2013finding} output multiple maps that potentially contain the symmetric map for a bilateral shape
but is hard to generalize to shapes with multiple symmetric axes. \cite{sahilliouglu2013coarse} considered the symmetry ambiguity when computing a single map to avoid symmetry flips. All of the above mentioned methods involve multiple maps, however, none of the them aim to recover \emph{all} possible symmetry-aware maps. The ones that produce multiple maps provide either a given class of maps (like the direct and the symmetric for bilateral shapes) or a set of maps without any property (like intermediate maps).

\subsection{Shape Matching with Functional Maps}
Our approach is based on the functional map representation~\cite{ovsjanikov2012functional}. The vast majority of methods that use this framework for shape matching start with a set of descriptor functions, derived from point signatures or from landmarks, and use them jointly with global map quality criteria to compute a correspondence ~\cite{kovnatsky2013coupled,aflalo2013spectral,nogneng17} (we refer to~\cite{ovsjanikov2017computing} for an overview). While computing \emph{a functional} map reduces to solving a least squares system, the conversion from a functional map to a point-wise map is not trivial and can lead to errors and noise~\cite{rodola-vmv15,ezuz2017deblurring}. 
To improve accuracy, several desirable map attributes have been promoted via regularizers for the functional map estimation first using geometric insights \cite{Eynard16,rodola2017partial,nogneng17,litany2017fully,burghard2017embedding,wang2018vector,ren2018continuous,wang2018kernel,gehre2018interactive,Shoham2019hierarchicalFmap}, and more recently using learning-based techniques \cite{FMNET,halimi2019unsupervised, roufosse2019unsupervised}. Nevertheless, despite significant progress, the reliance on descriptors and decoupling of continuous optimization and pointwise map conversion remains common to all existing methods.


\revised{
Our method is also related to spectral embedding methods. Some previous methods~\cite{jain2007,mateus08,rustamov07,ovsjanikov08} try to solve the shape matching (or symmetry detection) problem by directly aligning each dimension of the spectral embedding to obtain correspondences. In the functional map language this means forcing the functional maps to be diagonal~\cite{Nagar_2018_ECCV}. Our method aims to explore block-diagonal functional maps as initializations that we then refine to dense correspondences without imposing diagonality. As we demonstrate below, this is crucial to obtain high quality correspondences in diverse and challenging cases.
}

\subsection{Map Refinement}
A common strategy for improving estimated correspondences consists in iterative map refinement as a post-processing step, e.g.~\cite{solomon2016entropic,mandad2017variance,vestner2017product,vestner2017efficient}. The simplest refinement in the functional maps framework is the Iterative Closest Point algorithm in the spectral domain~\cite{ovsjanikov2012functional}. Recently, other more advanced refinement methods for both functional and pointwise maps have been proposed in \cite{ezuz2018reversible,ren2018continuous}, that, respectively, try to minimize the bi-directional geodesic Dirichlet energy, and promote the bijectivity, smoothness and coverage of the correspondences. When shape \emph{collections} are considered, a common strategy is to use cycle consistency constraints~\cite{wang2013,huang2014functional,wang2013exact}. Most closely related to ours, is a recent {\zo} method proposed in ~\cite{zoomout}, and based on iterative conversion between functional and pointwise maps. 

\paragraph{The graph isomorphism problem} 
We also mention briefly that computing all high quality (e.g., nearly isometric) maps between shapes is related to counting \emph{all isomorphisms} between graphs. This problem has been studied from a theoretical perspective, \cite{mathon1979note,kobler2012graph}, and although complexity equivalence results exist, no polynomial time algorithm is known.


\setlength{\tabcolsep}{0.52em}
\begin{table}[!t]
\caption{Our method outputs 8 maps for this pair of human shapes, where the first four maps contain the complete left-to-right, or back-to-front symmetries, and the last four contain a combination across the upper and lower body. We report map quality measures, including orthogonality ("Ortho"), Laplacian Commutativity ("lapComm"),  orientation preservation ("Orient"), edge length distortion ("EdgeDist"), and as-rigid-as-possible ("ARAP") distortion. Note that according to most metrics all maps are comparable.}
\label{tb:shape_pair_measure}\vspace{-3pt}
\footnotesize
\begin{tabular}{|c|c|c|c|c|c|c|c|c|}
\hline
\hspace{-5pt}\begin{overpic}[trim=5.5cm 1cm 6.5cm 2cm,clip,width=0.1\linewidth,grid=false]{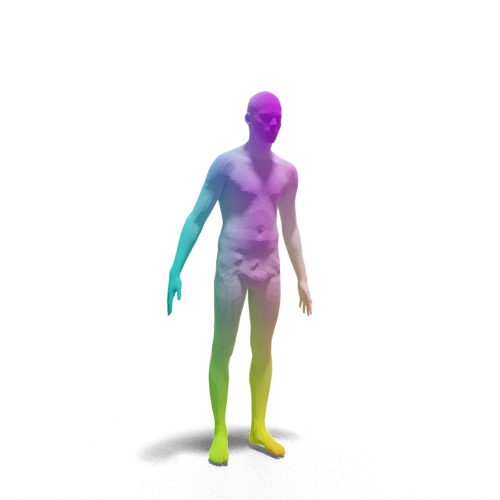}
\put(-8.5,90){\scriptsize (source)}
\end{overpic}\vspace{-0.5pt}\hspace{-5pt}
&
\hspace{-5pt}\begin{overpic}[trim=5.5cm 1cm 6.5cm 2cm,clip,width=0.1\linewidth,grid=false]{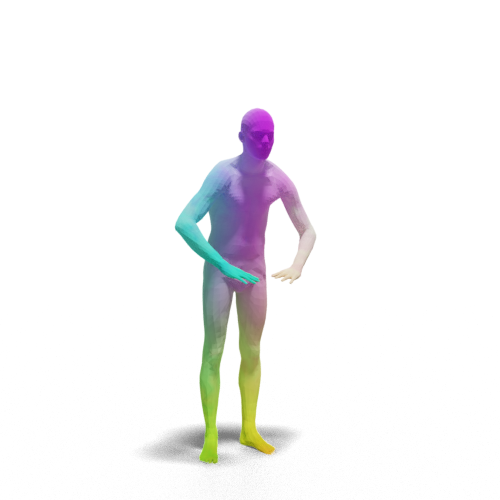}
\end{overpic}\vspace{-0.5pt}\hspace{-3pt}
& 
\hspace{-5pt}\begin{overpic}[trim=5.5cm 1cm 6.5cm 2cm,clip,width=0.1\linewidth,grid=false]{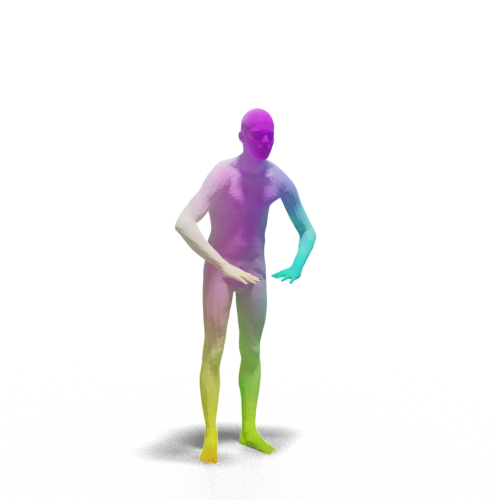}
\end{overpic}\vspace{-0.5pt}\hspace{-3pt}
&
\hspace{-5pt}\begin{overpic}[trim=5.5cm 1cm 6.5cm 2cm,clip,width=0.1\linewidth,grid=false]{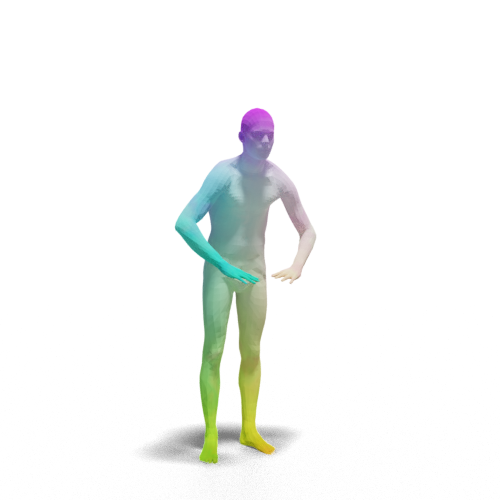}
\end{overpic}\vspace{-0.5pt}\hspace{-3pt}
&
\hspace{-5pt}\begin{overpic}[trim=5.5cm 1cm 6.5cm 2cm,clip,width=0.1\linewidth,grid=false]{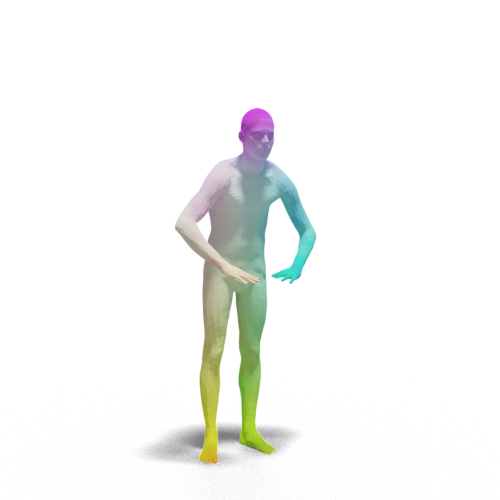}
\end{overpic}\vspace{-0.5pt}\hspace{-3pt}
&
\hspace{-5pt}\begin{overpic}[trim=5.5cm 1cm 6.5cm 2cm,clip,width=0.1\linewidth,grid=false]{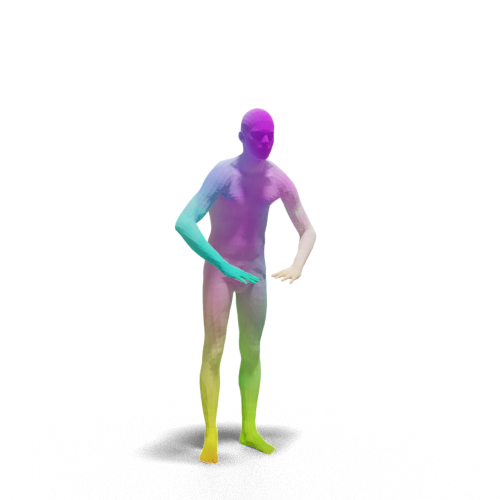}
\end{overpic}\vspace{-0.5pt}\hspace{-3pt}
&
\hspace{-5pt}\begin{overpic}[trim=5.5cm 1cm 6.5cm 2cm,clip,width=0.1\linewidth,grid=false]{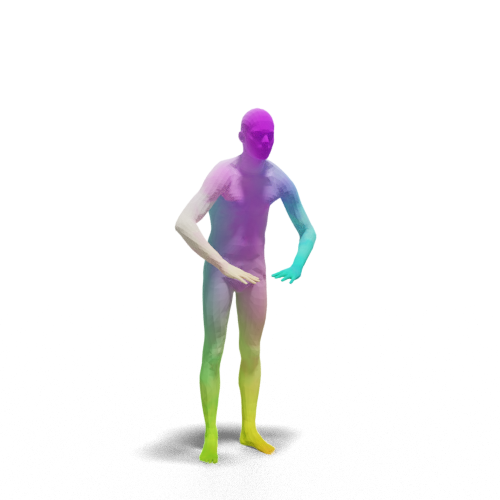}
\end{overpic}\vspace{-0.5pt}\hspace{-3pt}
&
\hspace{-5pt}\begin{overpic}[trim=5.5cm 1cm 6.5cm 2cm,clip,width=0.1\linewidth,grid=false]{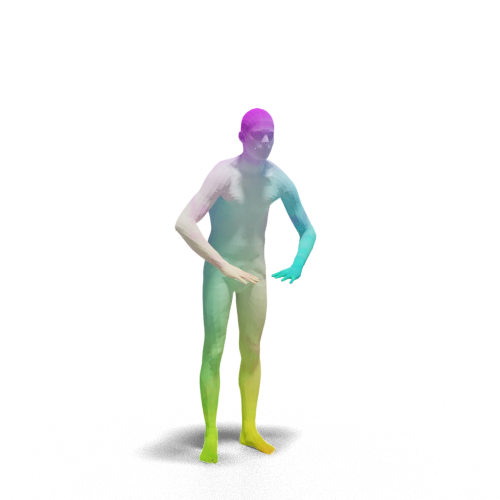}
\end{overpic}\vspace{-0.5pt}\hspace{-3pt}
&
\hspace{-5pt}\begin{overpic}[trim=5.5cm 1cm 6.5cm 2cm,clip,width=0.1\linewidth,grid=false]{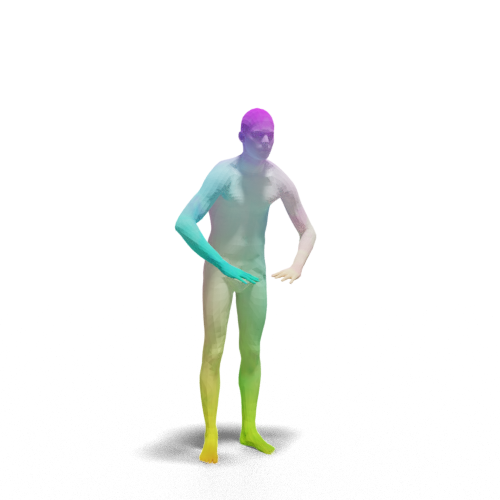}
\end{overpic}\vspace{-0.5pt}\hspace{-3pt}
\\
Measure & map1 & map2 & map3 & map4 & map5 & map6 & map7 & map8 \\ \hline
Ortho & 19.9 & 20.3 & 19.9 & 19.9 & 18.6 & 17.3 & 19.3 & 20.5 \\ \hline
lapComm & 12.0 & 11.3 & 16.1 & 16.7 & 21.2 & 21.6 & 23.0 & 24.3 \\ \hline
Orient & 0.72 & 1.24 & 1.28 & 0.85 & 0.79 & 1.23 & 0.91 & 1.18 \\ \hline
EdgeDist & 1.46 & 1.97 & 2.56 & 3.73 & 4.54 & 4.30 & 5.73 & 4.57 \\ \hline
ARAP & 1.28 & 1.39 & 1.47 & 1.45 &2.99 & 2.90 & 3.36 & 3.36 \\ \hline
\end{tabular}
\end{table}

\section{ Motivation, Background \& Notation}
Our goal is to find multiple high-quality maps on a shape pair or a single shape that  encode different symmetries. In order to motivate our work, we would like to briefly review three simple solutions to the problem.

The first idea is to exhaustively enumerate all maps in the point-wise map space (with size $O(n^n)$, $n$ is the number of vertices) and select some maps based on some criterion (e.g., geodesic distortion energy). This is infeasible when the shapes have more than about 10 vertices due to the size of the space.

The second idea would be to use random sampling to only evaluate a subset of the maps. However, the probability of obtaining a high-quality map from random sampling is almost zero given the large size of the search space.

A third idea would be to take an existing optimization algorithm and start it with multiple different initializations. 
Common approaches for initializing maps include using pointwise descriptors, e.g., \cite{vestner2017product,eisenberger2019divergence} or initial segment or landmark correspondences \cite{ren2018continuous,ezuz2018reversible,kleiman2018robust}. For these methods it is actually very challenging to modify the initialization in a meaningful way so that the optimization methods would explore different parts of the map space. It would be necessary to run these methods with different random initializations. 
However, there is also no guarantee that different parts of the space will be explored and that a random initialization will lead to a meaningful map. While this approach is more promising than randomly sampling maps, it would still require a large number of random initializations. We will compare to this approach as a baseline to demonstrate that only few distinct maps can be recovered compared to our method (e.g. Table~\ref{tb:res:selfSymm})


A second challenge also becomes evident. Simply proposing a set of maps is not sufficient for an application. We need to have a map selection algorithm that either selects the direct map, a particular symmetric map, or all meaningful symmetric maps from a set of candidate maps.


We propose a novel  algorithm, called \emph{MapTree}, that tackles these two challenges. First of all, instead of dealing with the unorganized point-wise map search space, we propose to solve the same problem in the \emph{spectral} domain based on the following observations. (1) each point-wise map can be represented as a functional map in a reduced basis~\cite{ovsjanikov2012functional}. (2) the functional-map representation utilizes the eigenfunctions associated with specific frequencies (ordered eigenvalues). Therefore, the functional-map space can be well organized w.r.t. the frequencies which makes it much easier for map exploration (3) a single point-wise map can be converted to functional maps with different dimensions, where the low dimension encodes the global information of the point-wise map (such as the symmetry orientation), while the high dimension encodes the local details of the map. Therefore, to disambiguate two maps with good-quality, we can convert them to functional maps with different dimension for comparison.

Based on these observations, we propose a tree structure to explore the functional map space, where at low dimensions we can enumerate all possible functional maps in good-quality.
The idea of exploring low-dimensional functional maps has been used for symmetry detection in e.g.,~\cite{Nagar_2018_ECCV} and to some extent in~\cite{wang2017group}. However, those methods are limited by specific choice of the dimensionality (typically very low such as 6-10) in order to maintain efficiency. Moreover, our method is not limited to symmetry detection and can naturally be applied to compute multiple maps across shapes.

Here we clarify the notations that are used in our method. Given two shapes $S_1$ and $S_2$ represented as triangle meshes, we associate to each the Laplace-Beltrami operator using the cotangent weight discretization \cite{meyer03}. The first $k$ eigenfunctions of this operator on shape $S_i$ form a basis denoted by $\Phi_{S_i}^{(k)} = \big[\phi_1^{S_i}, \phi_2^{S_i},\cdots, \phi_k^{S_i}\big]$, having the eigenfunctions stored as columns in a matrix.

Given a pointwise map $T_{12}: S_1 \rightarrow S_2$, that maps each vertex in shape $S_1$ to a vertex in shape $S_2$, we represent it with a binary matrix $\Pi_{12}$, such that the $\Pi_{12}(i, j) = 1$ if $T_{12}(i) = j$ and 0 otherwise. Throughout our paper, we use $\Pi$ to denote pointwise correspondences implicitly restricting the matrix to be binary with exactly one value $1$ at each row. 

As introduced in ~\cite{ovsjanikov2012functional} the functional map representation of $T_{12}$ is a linear transformation mapping \emph{functions} on $S_2$ to \emph{functions} on $S_1$  (note the change in direction compared to $T_{12}$). When a functional map corresponds to a pointwise map it can be written as a matrix:

\vspace{-2mm}
\begin{equation}\label{eq:mtd:pmap_to_fmap}
C_{21} = \Phi_{S_1}^{\dagger}\Pi_{12}\Phi_{S_2},
\end{equation}
where $\dagger$ denotes the Moore-Penrose pseudo-inverse.

\section{Map Space Exploration\label{sec:map_space}}
Our overall objective is to recover a set of high quality maps between a given shape pair.  As mentioned above, the recovered maps should be sufficiently different and, ideally all be approximately smooth and near isometric. 

To tackle this problem, our main idea is to exploit the spectral map representation, in order to efficiently enumerate accurate low-di\-men\-sio\-nal functional maps, which we refine to near-isometric pointwise correspondences. We organize the functional maps into a tree structure, where each level corresponds to the map size, and a node corresponds to all functional maps that share a particular leading principal submatrix. Our general approach then consists in progressively expanding this tree to obtain high quality diverse solutions. In this section, we first discuss how the spectral techniques can help us explore the map space. We then describe our approach for recovering the map tree hierarchy starting with a pair of shapes.

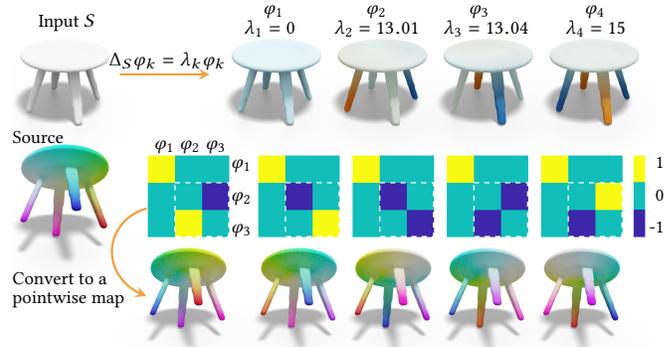
\begin{figure}[!t]
    \centering
    \input{figures/eg_table.tex}
    \vspace{-20pt}
    \caption{Illustration of repeated eigenvalues. \emph{Top}: A table with four rotational and four reflectional symmetries, and the first four eigenvalues/functions of its LB operator. Note that the second ($\phi_2$) and the third eigenfunctions ($\phi_3$) correspond to similar eigenvalues. \emph{Middle and Bottom}: we plot different functional maps and associated pointwise correspondences via color transfer. Note that that different permutations between $\phi_2$ and $\phi_3$ lead to different self-symmetric maps.}
    \label{fig:mtd:eg_table}
\end{figure}

\subsection{How to efficiently explore the space of sufficiently different but high-quality maps?}
As noted above, Fig.~\ref{fig:landscape} illustrates that the commonly used geodesic distortion energy is strongly non-convex with multiple strong local minima. Fig.~\ref{fig:contour} further shows that two maps with similar geodesic distortion error can encode completely different symmetries. Finally, in Table~\ref{tb:shape_pair_measure} we show another shape pair, and report multiple commonly-used map quality criteria. Remark that these criteria cannot disambiguate maps with different symmetry effectively.

To better structure the space of different maps, we propose to use the multi-scale functional map representation. Namely, for each pointwise map $T_{12}$ from $S_1$ to $S_2$, we can convert it into a functional map $C_{21}$ by Eq.~\eqref{eq:mtd:pmap_to_fmap} of a particular size. Our general observation is that small (low-dimensional) functional maps provide a convenient way of structuring the map space for three reasons: 1) they allow to encode sufficiently different pointwise maps efficiently 2) as demonstrated below they can be explored effectively due to their small size, 3) they can be \emph{refined} to obtain accurate pointwise correspondences.

The discriminating ability of low-dimensional functional maps comes from two complementary results. First, given two pointwise maps that are sufficiently \emph{close} to each other, their functional map representation, \emph{especially in the low frequencies} will be similar. Indeed, we have the following theorem (proved in Appendix~\ref{append:proof}):

\begin{theorem}\label{thm:bound}
 Given two pointwise maps $T_{12}^1, T_{12}^2: S_1 \rightarrow S_2$ that are $\delta$-close, their functional map matrix representation $C_{21}^1, C_{21}^2$ will satisfy the following bound $\sum_j \big(C_{21}^1(j,i) - C_{21}^2(j,i)\big)^2 < (\delta c \lambda_i)^2 A_1\;\forall i$.
\end{theorem}
\noindent Here two maps $T^1, T^2$ are $\delta$-close if for any vertex $v$ the geodesic distance between $T_{12}^1(v)$ and $T_{12}^2(v)$ is smaller than $\delta$, $c$ is some constant (which depends on $S_2$), $A_1$ is the surface area of shape $S_1$, and $\lambda_i$ is the $i$-th eigenvalue of shape $S_2$. This bound suggests that for two maps that are close to each other, i.e., $\delta$ is small, we need larger $i$ (higher frequencies) to disambiguate two maps. 

Moreover, while Theorem~\ref{thm:bound} only establishes the bound in one direction, showing that nearby pointwise maps will result in similar low frequency functional maps, a complementary result can also be found. Specifically if two pointwise maps are \emph{sufficiently different} their low frequency functional map representations must also be different. Intuitively, this is because for two different pointwise maps it must be possible to find indicator functions of regions that are mapped differently by them. If those regions are large enough so that their projection onto the low frequency eigenfunctions are significant, this implies that the functional maps must be different in the sense of the Frobenius norm. We formalize this intuition in Appendix \ref{appendix:lower_bound}.

In summary, there are several advantages of the spectral representation via functional map: (1) the Frobenius-norm between two functional maps is a cheap but relatively accurate approximation of the distance between two pointwise maps. (2) for two maps that are far from each other (e.g., with different symmetry), we only need relatively low frequencies to convert functional maps that can disambiguate the two pointwise maps (see Fig.~\ref{fig:mtd:eg_table} for an example). (3) for two maps that are close to each other, we then need higher frequencies for disambiguation. Therefore, we can organize pointwise maps in the spectral domain along the frequencies. 

To exploit these properties we propose a pipeline that effectively explores low-frequency functional maps and then refines them to higher-frequency functional maps and accurate pointwise correspondences. For the latter stage, we extend the approach proposed in~\cite{zoomout} that allows to obtain high quality maps from small low frequency functional maps.

\begin{algorithm}[!t]
\DontPrintSemicolon
\SetKwData{Left}{left}\SetKwData{This}{this}\SetKwData{Up}{up}
\SetKwFunction{Union}{Union}\SetKwFunction{FindCompress}{FindCompress}
\SetKwInOut{Input}{Input}\SetKwInOut{Output}{Output}
\Input{Two shapes $S_i$ (i = 1,2) with corresponding eigenfunctions $\{\phi_k^{S_i}\}$ and eigenvalues $\{\lambda_k^{S_i}\}$; an initial functional map $C_{\text{ini}}\in \mathbb{R}^{k_2\times k_1}$; A refinement method $\mathcal{R}$
}
\Output{A set of functional maps $\bm{\mathcal{C}}$ expanded from $C_{\text{ini}}$}
(1) group similar eigenvalues for both shapes: find the index $p_i$ for shape $S_i$ such that \revised{$\lambda_{p_i}^{S_i} - \lambda_{k_i}^{S_i} \le \epsilon < \lambda_{p_i + 1}^{S_i} - \lambda_{k_i}^{S_i}, i = 1,2$}\;
(2) enumerate the mappings between $\{\phi_{k_1}^{S_1},\cdots,\phi_{p_1}^{S_1}\}$ and $\{\phi_{k_2}^{S_2},\cdots,\phi_{p_2}^{S_2}\}$; store the assignment as a $p_2\times p_1$ matrix $C_{\text{perm}}$ (with entry $0,\pm 1 $) \;
(3) construct $C = \begin{pmatrix}C_{\text{ini}} & \mathbb{0}\\
\mathbb{0} & C_{\text{perm}}\end{pmatrix}$, where $\mathbb{0}$ is a zero-matrix\;
(4) refine each $C$ using the given refinement method $\mathcal{R}$. Convert the refined maps to functional maps of size equal to that of $C$, and output the result.
\caption[caption]{Functional Map Expansion Rules}
\label{alg:mtd:enumeration}
\end{algorithm}

\subsection{Hierarchical Map Recovery}
 
 As suggested above, we propose to organize the functional maps in a tree structure (which we call the Map Tree) from low frequency to high frequency. The nodes at each level $k$ of the tree represent different functional maps of size $k$-by-$k$, and two nodes across adjacent levels $k$ and $k+1$ are connected if and only if the functional map matrices have the same leading principal $k$-by-$k$ submatrix. Such a map tree structure provides a convenient way of map organization using the ordered nature of the Laplacian spectrum. It also helps to explore the map space efficiently.
 


Our main idea is to build the map tree in a coarse to fine fashion. Namely, we start with the root node and progressively test whether the descendant nodes can arise from a high quality map. Our basic algorithm, which we call $\textbf{Algorithm A},$ is summarized as follows:
\begin{enumerate}[leftmargin=*]
\item Initialize the set $\mathcal{S} = \{C_0\}$ to contain the single functional map of size $1 \times 1$ equal to the ratio of the values of the constant basis functions.
\item For each $C_i \in \mathcal{S}$:
  \begin{enumerate}
  \item Let $C_i^{+}$ and $C_i^{-}$ be two functional maps obtained by adding an extra row and column to $C_i$, with zero values, except for the bottom right diagonal, set to $1$ and $-1$ respectively.
  \item Apply a map refinement algorithm $\mathcal{R}$ given $C_i^{+}$ as initialization. If the resulting pointwise map $T$ is of sufficiently high quality, convert $T$ to a functional map of the same size as $C_i^{+}$ and add it to $\mathcal{S}$. Do the same for $C_i^{-}$. 
  \item Repeat the previous two steps until all functional maps in $\mathcal{S}$ have size greater than some threshold $\kappa$.
  \end{enumerate}
\end{enumerate}

\begin{algorithm}[!t]
\DontPrintSemicolon
\SetKwData{Left}{left}\SetKwData{This}{this}\SetKwData{Up}{up}
\SetKwFunction{Union}{Union}\SetKwFunction{FindCompress}{FindCompress}
\SetKwInOut{Input}{Input}\SetKwInOut{Output}{Output}
\Input{A pair of shapes $S_i$ (i = 1,2) with corresponding eigenfunctions $\{\phi_k^{S_i}\}$ and eigenvalues $\{\lambda_k^{S_i}\}$; some stopping criterion and some tree pruning rules}
\Output{A set of maps and the corresponding map tree}
\textbf{Initialization} : Initialize the map tree with a node $C_{\text{ini}} = \frac{\sum \phi_{1}^{S_2}}{\sum \phi_{1}^{S_1}}$ with the status "un-explored" \;
 \While{Stopping Criterion Not Met}{
\For{each \textbf{un-explored leaf} $C^*$ of the map tree}{
    (1) Apply Algo.~\ref{alg:mtd:enumeration} and obtain a set of functional maps $\bm{\mathcal{C}}$ \;
    (2) Update the map tree by attaching each $C\in\bm{\mathcal{C}}$ to the node $C^*$ with the default status "un-explored"\;
    (3) Prune the tree w.r.t. the given pruning rule and label the pruned leaves as "explored".
}
  }    
\caption[caption]{Map Tree Exploration}
\label{alg:mtd:mapTree}
\end{algorithm}  

Intuitively, this algorithm progressively tries different diagonal functional map initializations and applies a refinement algorithm $\mathcal{R}$ to them. In order to analyze this map exploration approach, we need an assumption on the refinement algorithm. Specifically, we will call algorithm $\mathcal{R}$ \emph{complete} if, given an initial functional map $C$ of size $k \times k$, and assuming there exists at least one isometric map, whose functional map representation has $C$ as the leading principal $k \times k$ sub-matrix, then algorithm $\mathcal{R}$ will recover such an isometry, and an empty set otherwise.
  
  The following theorem shows that under certain assumptions, $\textbf{Algorithm A}$ is guaranteed to recover the map tree up to level $\kappa$.

  \vspace{-1mm}
  \begin{theorem}
    \label{thm:map_tree}
    Suppose the Laplacians of $S_1, S_2$ have the same eigenvalues, none of which are repeating. Let $\mathcal{M}$ denote the tree constructed from \emph{all isometries} between $S_2 \rightarrow S_1$. Further suppose that the refinement algorithm $\mathcal{R}$ is \emph{complete}. Then the subtree of $\mathcal{M}$ until level $\kappa$ will coincide with the tree built from the output of Algorithm A.
  \end{theorem}

  \vspace{-4mm}
  \begin{proof}
    See Appendix.
  \end{proof}

We remark that although there are $2^k$ possible functional maps of size $k \times k$ that have $+1$ or $-1$ on the diagonal the complexity of \textbf{Algorithm A} depends on the number of recovered isometries. For example if there exists only one isometric correspondence, this algorithm will only test $2k$ initializations.

\paragraph{Practical Implementation}
\revised{
We propose the Map Tree Exploration algorithm (Algo.~\ref{alg:mtd:mapTree}) with Functional Map Expansion Rules (Algo.~\ref{alg:mtd:enumeration}) to implement the idea of Algorithm A with several practical modifications.
}

First, we take special care in the presence of eigenvalues that are close to being repeating, which can arise particularly for symmetric shapes (see  Fig.~\ref{fig:mtd:eg_table} for an example). In this case, as has been observed in, e.g. \cite{ovsjanikov08,wang2017group} the corresponding eigenfunctions will form an isometry-invariant \emph{subspace}. Therefore, we must allow not only $\pm 1$ on the diagonal, but also consider all orthogonal \emph{combinations} among the corresponding eigenfunctions. To achieve this, we need two building blocks: (1) detecting repeating eigenvalues, and (2) enumerating combinations among the grouped eigenfunctions. Importantly, since our method tries a set of initializations for refinement, we only need a \emph{conservative} estimate for declaring eigenvalues to be repeating, since poor initalizations will be discarded. Therefore, we detect repeating eigenvalues by simple thresholding. For enumerating possible initializations we consider all permutations (allowing sign flips) among grouped eigenfunctions. We have observed this to perform remarkably well, due, in particular, to the robustness of our refinement method, although theoretically the invariant subspace can be continuous. Algorithm~\ref{alg:mtd:enumeration} describes the details of the functional map expansion rules used in practice. We stress that our approach is geared towards a \emph{discrete} set of maps, and we do not consider continuous spaces, such as isometries between two spheres. \revised{Also note that the detected repeating eigenvalues of two shapes might not have the same size, therefore, the expanded functional map after applying Algorithm~\ref{alg:mtd:enumeration} can be rectangular.}

Algorithm~\ref{alg:mtd:mapTree} describes our complete map tree construction method, which uses the expansion Algorithm \ref{alg:mtd:enumeration} as a subroutine. Note that Algorithm~\ref{alg:mtd:mapTree} also relies on pruning rules for deciding when a map is of sufficiently good quality. In practice we use the following:
\begin{enumerate}[leftmargin=*]
    \item Discard a refined map if its orthogonality error $ E_{\text{ortho}} \big(C\big) > \epsilon_1$ or its Laplacian Commutativity error $E_{\text{lapComm}}\big(C\big) > \epsilon_2$.
    \item Prune the leaf $C$ that converges to the same functional map as another leaf.
\end{enumerate}

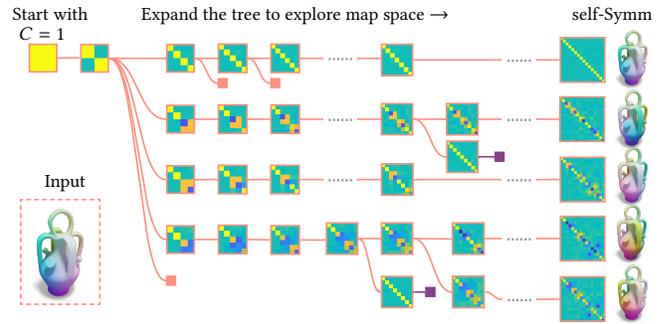
\begin{figure}[!t]
    \centering
    \begin{overpic}[trim=0cm 0cm 0cm 0cm,clip,width=1\linewidth,grid=false]{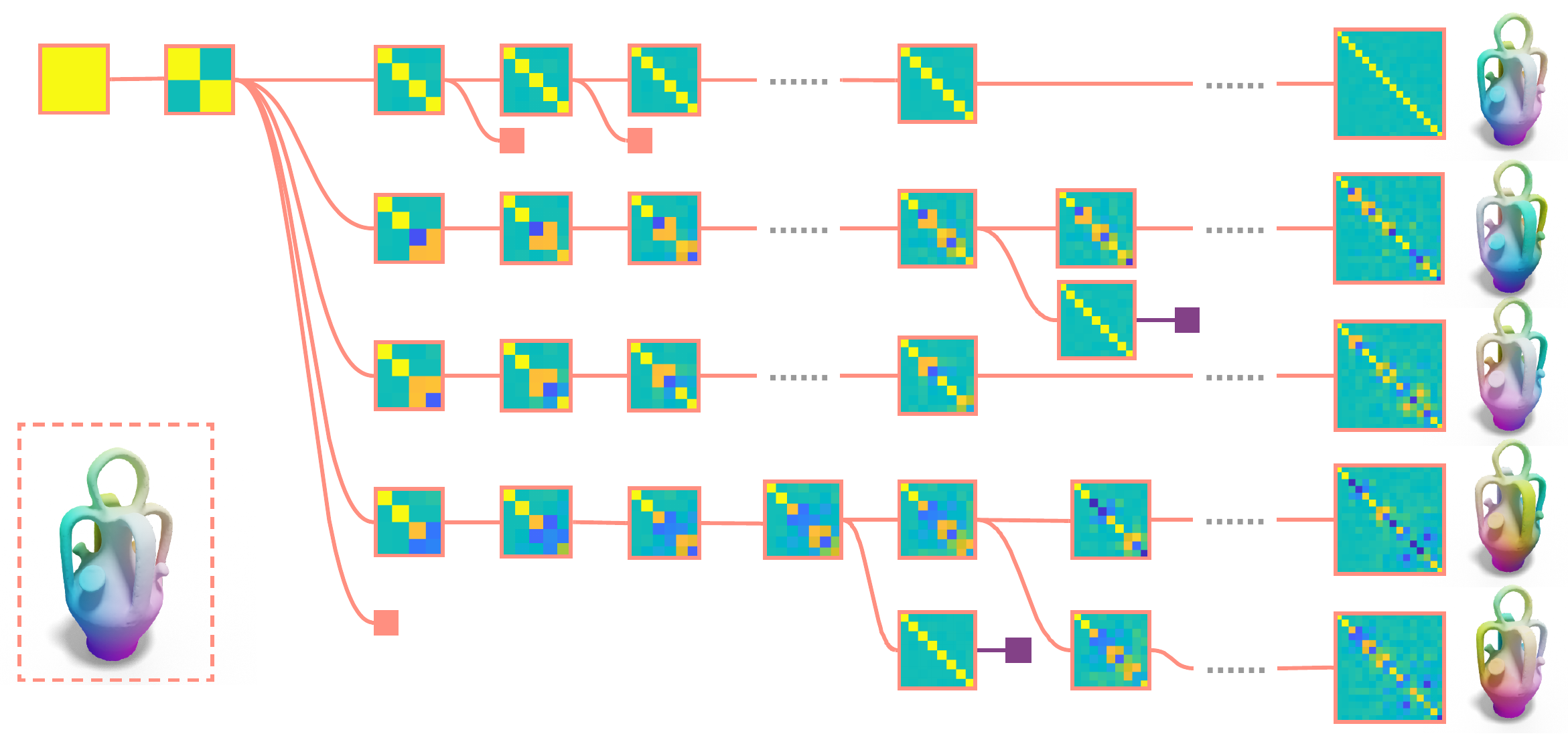}
    \put(5,22){\footnotesize Input}
    \put(0,48){\footnotesize Start with}
    \put(1,45){\footnotesize $C = 1$}
    \put(20,48){\footnotesize Expand the tree to explore map space $\rightarrow$}
    \put(87,48){\footnotesize self-Symm}
    \end{overpic}
    \vspace{-20pt}
        \caption{Illustration of Algorithm~\ref{alg:mtd:mapTree} for exploring the self-symmetry space of a vase. The red/purple squares represent nodes pruned due to poor quality and duplicates of refined maps respectively. The last column shows the final computed functional and corresponding point-wise maps that capture different self-symmetries. \vspace{-3mm}}
    \label{fig:mtd:eg_maptree_vase}
\end{figure}

Since the orthogonality ($E_{\text{ortho}}\big(C\big) = \Vert C C^T - I\Vert_F^2$) and the Laplacian Commutativity ($E_{\text{lapComm}}\big(C\big) = \Vert C\Delta_1 - \Delta_2 C\Vert_F^2$) are strong indicators for detecting bad functional maps, We use them to remove the leaves/functional maps that are unlikely to be good. 
We introduce rule (2) to avoid initializations that converge to the same functional map after refinement. 
The rationale behind it is that, as we show in the next section, our new refinement method has strong convergence power, capable of producing high quality maps even from poor initializations. Thus, we use rule (2) to remove initializations that lead to the same maps to avoid redundancy. Fig.~\ref{fig:mtd:eg_maptree_vase} shows how a map tree is constructed using our algorithm. 

the

\subsection{Bijective \zo}
In Algorithm~\ref{alg:mtd:enumeration} (step 4), we need a refinement method $\mathcal{R}$ to refine the expanded functional map from the enumeration. For efficiency, we can use some spectral-based methods such as ICP~\cite{ovsjanikov2012functional} or \zo~\cite{zoomout}. \zo~ is the current state-of-the-art refinement for near-isometric shape matching, which progressively increases the frequency during the refinement. Specifically, it proposes the following energy to minimize, \revised{where each principal $k\times k$ submatrix is constrained to be orthonormal}:
\begin{equation}\begin{split}
    E_{\text{ZM}}(C_{21}) =& \sum_{k} \frac{1}{k}\big\Vert C_{21}^{(k)}(C_{21}^{(k)})^T - I_k  \big\Vert^2 \\
    & \exists \Pi_{12}, \text{s.t.} \;  C_{21} = \Phi_{S_1}^\dagger \Pi_{12} \Phi_{S_2}
\end{split}\end{equation}
 To remove the direction bias and improve upon \zo, we propose to add the bijectivity constraint on the two functional maps mapping in different directions. Specifically, we propose to minimize the following energy:
\begin{equation}\begin{split}
    E(C_{12}, C_{21}) &= E_{\text{ZM}}(C_{12}) + E_{\text{ZM}}(C_{21}) + \\
    & \sum_{k}\frac{1}{k}\big\Vert C_{12}^{(k)}C_{21}^{(k)} - I_k\big\Vert^2 + 
    \sum_{k}\frac{1}{k}\big\Vert C_{21}^{(k)}C_{12}^{(k)} - I_k\big\Vert^2 \\
 & \exists \Pi_{12}, \Pi_{21}, \text{s.t.} \;  C_{12} = \Phi_{S_2}^\dagger \Pi_{21} \Phi_{S_1}, C_{21} = \Phi_{S_1}^\dagger \Pi_{12} \Phi_{S_2}
\end{split}\end{equation}
We use a similar half-quadratic splitting technique as~\cite{zoomout} to decouple the pointwise maps $\Pi$ and the corresponding functional maps $C$. We provide the pseudo-code and implementation details in the appendix.

\setlength{\tabcolsep}{0.36em}
\begin{table}[!t]
\caption{\revised{We compare our MapTree method to other simple baselines on a pair of human shapes from the FAUST dataset. We sample 500 random point-wise maps and apply different refinement methods. We then report the quality of the refinement methods and the total runtime.}}
\label{tb:res:randIni}\vspace{-3pt}
\footnotesize
\begin{tabular}{|l|c|c|c|c|c|c|}
\hline
Methods & \# diffMaps & Acc & GeoDist & \revised{Dirichlet} & \revised{Conformal} & runtime (h) \\ \hline\hline
Random Ini & 500 & 0.623 & 2.088 & 1092 & 6.958 & - \\ 
ICP$_{50}$ & 500 & 0.463 & 1.880 & 75.78 & 21.83 & 1.946 \\
PMF$_{50}$ & 127 & 0.050 & \textbf{0.286} & 44.07 & 4.049 & 82.21 \\
BCICP$_{50}$ & 500 & 0.154 & 1.727 & 30.49 & 13.83 & 63.62 \\
RHM$_{50}$ & 500 & 0.285 & 1.726 & 53.76 & 6.956 & 5.841 \\
ZoomOut$_{20\cdots 50}$ & 500 & 0.372 & 1.767 & 48.09 & 8.697 & 0.221 \\
ZoomOut$_{2\cdots20}$ & 454 & 0.031 & 0.775 & 13.68 & 2.944 & 0.080 \\ \hline\hline
MapTree (Ours) & \textbf{10} & \textbf{0.019} & 0.375 & \textbf{11.26} & \textbf{1.815} & \textbf{0.002} \\ \hline
\end{tabular}
\end{table}

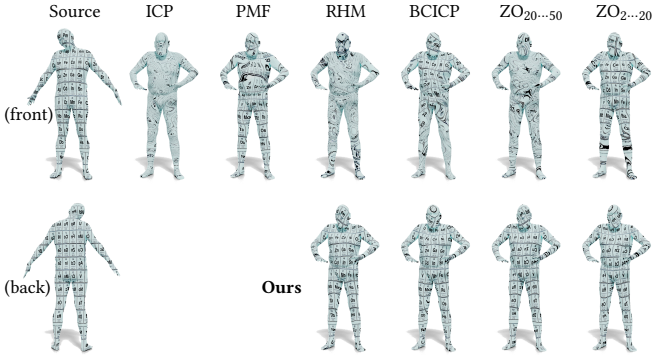
\begin{figure}[!t]  
    \centering
    \begin{overpic}[trim=0cm 0cm 0.5cm 0cm,clip,width=1\linewidth,grid=false]{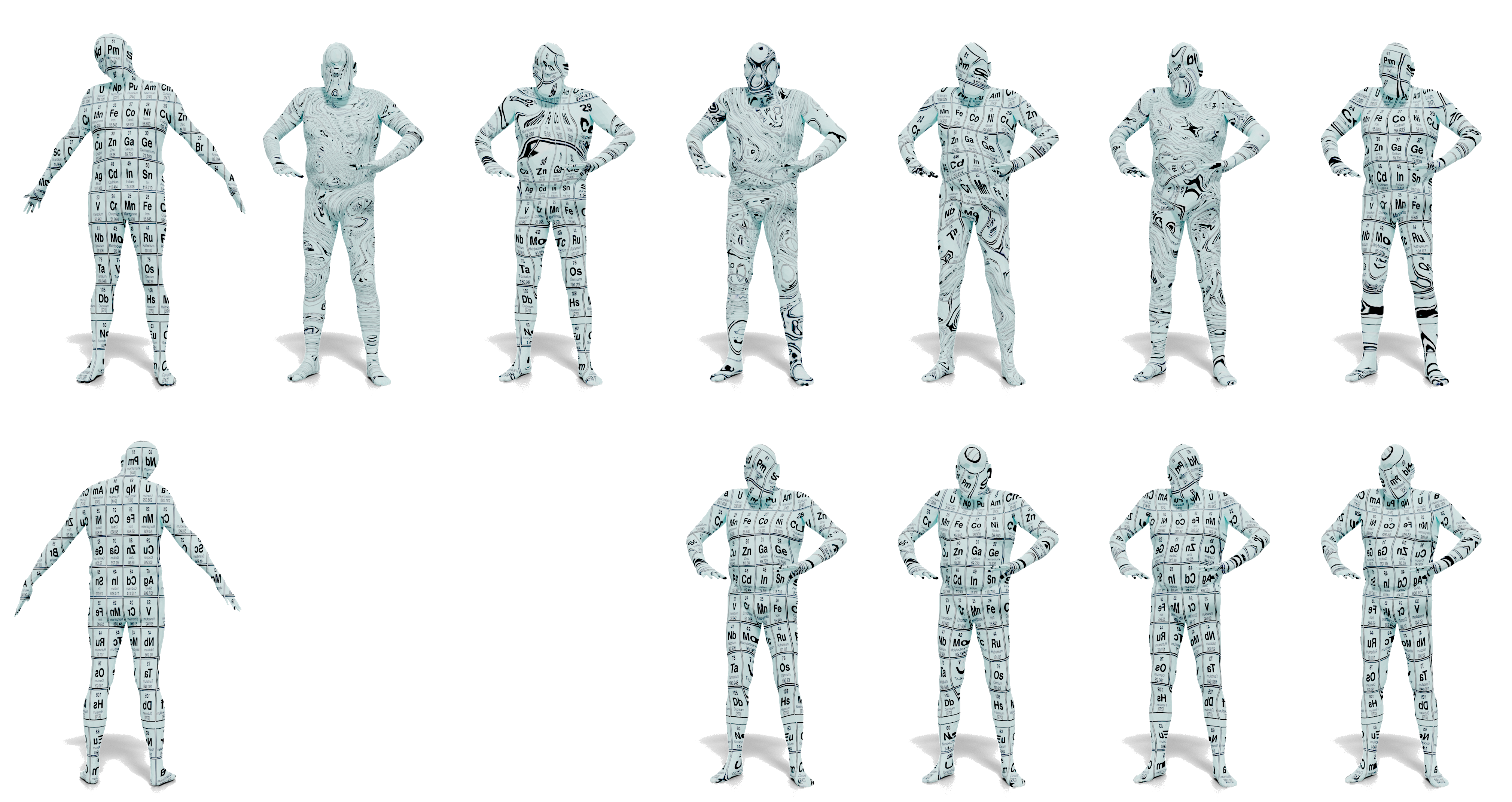}
    \put(-2,37){\footnotesize (front)}
    \put(-2,10){\footnotesize (back)}
    \put(5,53){\footnotesize Source}
    \put(20,53){\footnotesize ICP}
    \put(34,53){\footnotesize PMF}
    \put(48,53){\footnotesize RHM}
    \put(61,53){\footnotesize BCICP}
    \put(75,53){\footnotesize ZO$_{20\cdots50}$}
    \put(90,53){\footnotesize ZO$_{2\cdots20}$}
    \put(38,10){\footnotesize \textbf{Ours}}
    \end{overpic}
    \vspace{-20pt}
    \caption{
    \revised{
    We show the shape pair that is tested in Table~\ref{tb:res:randIni}. \emph{Top}: we show the best refined map of each method. \emph{Bottom}: we show the four obtained maps using our method, including the direct map, symmetric map, back-to-front map, and doubly flipped map via texture transfer.}}
    \label{fig:eg_faust_texture}
\end{figure}\vspace{-3pt}

\section{Results}
In this section, we show extensive quantitative and qualitative results on three applications: (1) multi-solution shape matching, where our method is applied to produce multiple maps/symmetries; (2) standard single-solution non-rigid shape matching;
(3) self-symmetry detection.
In addition, we provide information on the parameters of the algorithm and its runtime. 

To compare different methods, we \revised{measure the accuracy and the smoothness of the computed maps}:
\begin{itemize}[leftmargin=*]
    \item \textbf{Accuracy}: The \emph{accuracy} of a given map is obtained by measuring the \emph{geodesic} distance between the computed correspondence and the given ground-truth correspondence, where the geodesic distance is normalized by the surface area to make this measurement comparable across different shape pairs.
    \item \revised{
    \textbf{Smoothness}: We use three different metrics to measure the smoothness of a given map, namely, the geodesic distortion, Dirichlet energy, and conformal distortion:
    \begin{itemize}
        \item \textbf{Geodesic Distortion}: Assume that matrix $G_k(i,j)$ stores the normalized geodesic distances between all pairs of vertices $i$ and $j$ for shape $S_k$. Then for a given map $T_{12}: S_1 \rightarrow S_2$, we measure its \emph{geodesic distortion} on every pair of $(i, j)$ as:  $$ \sum_{i}\sum_{j} \Big(G_1(i,j) - G_2\big(T_{12}(i), T_{12}(j)\big)\Big)^2 $$
        \item \textbf{Dirichlet Energy}: We measure the Dirichlet energy of the \emph{mapped} vertex positions on the target shape. 
        \item \textbf{Conformal Distortion}: we use the same metric as proposed in{~\cite{ezuz2018reversible}}, which is modified from~{\cite{hormann2000mips}}: for each face $f$ in $S_1$, we compute $\frac{\sigma_1}{\sigma_2}+\frac{\sigma_2}{\sigma_1} - 2$, where $\sigma_i$ are the singular values of the linear transformation that maps $f$ from $S_1$ to $S_2$. We then average this value over all the faces in $S_1$.
    \end{itemize}
    }
\end{itemize}

\subsection{Multi-Solution Shape Matching}

Our algorithm MapTree has no published direct competitor for the problem of multi-solution shape matching.
We therefore try to create a baseline method that consists of a modification of existing techniques. The most suitable idea seems to be to start with many different random point-to-point maps as initialization and use an existing optimization technique.
Specifically, we randomly generated 500 pointwise maps between a human shape from the FAUST-remeshed dataset, and apply different refinement methods including recent state-of-the-art methods RHM~\cite{ezuz2018reversible} and ZoomOut~\cite{zoomout}. 
A suitable competitor must be able to at least find the 
symmetric or the direct map in such an example. Therefore, when measuring the map accuracy, we consider both the direct and symmetric direction and pick the one that gives a smaller error. Accuracy and \revised{different smoothness metrics} are reported in Table~\ref{tb:res:randIni}.

As a conclusion, we can observe that only {\zo} works in combination with random initialization. Other methods have an excessive computation time (e.g. over 50h for PMF and BCICP) and with the exceptions of PMF they cannot converge to competitive high-quality maps for any of the 500 random initializations. See Fig.~\ref{fig:eg_faust_texture} for a visual comparison of the best map found by each method.
We thus select random initialization + {\zo} as baseline method.

\setlength{\tabcolsep}{0.48em}
\begin{table}[!t]
\caption{Quantitative comparison to (1) RandIni + {\zo}, (2) IntSymm, and (3) OrientRev + \zo. We report the number of detected high-quality maps (i.e., corresponding to some intrinsic symmetries) and the runtime on several shapes with different resolution.}\label{tb:res:selfSymm}\vspace{-3pt}
\footnotesize
\begin{tabular}{|c|c|c|cccc|cccc|}
\hline
\multirow{2}{*}{Shape} & \multirow{2}{*}{\begin{tabular}[c]{@{}l@{}}Figure \\ Number\end{tabular}} & \multirow{2}{*}{\#Vtx} & \multicolumn{4}{c|}{\# High-quality maps} & \multicolumn{4}{c|}{Runtime (sec)} \\ \cline{4-11} 
 &  &  & (1) & (2) & (3) & Ours & (1) & (2) & (3) & Ours \\ \hline\hline
Table & Fig.~\ref{fig:mtd:eg_table} & 14K & 1 & 1 & 1 & \textbf{7} & 2359 & 8.63 & 13.0 & 39.0 \\
Vase & Fig.~\ref{fig:mtd:eg_maptree_vase} & 15K & 2 & 1 & 1 & \textbf{4} & 1683 & 4.4 & 11.6 & 25.2 \\
Cup & Fig.~\ref{fig:mtd:eg_cup} & 30K & \textbf{3} &1 & 1 & \textbf{3}& 9474 & 19.8 & 34.6 & 30.0 \\
Glasses & Fig.~\ref{fig:mtd:eg_glasses} & 2K & 1 & 1 & 1 & \textbf{3} & 44.61 & 0.82 & 1.71 & 3.24 \\
Knot & Fig.~\ref{fig:res:knot} & 5K & 3 & 0 & 1 & \textbf{44} & 350.6 & - & 3.23 &  56.2\\
Legs & Fig.~\ref{fig:res:garment} & 12K & 1 &1 & 1 & \textbf{3} & 2016 & 2.15 & 11.2 & 3.69 \\
Baby & Fig.~\ref{fig:res:eg_angle} & 2K & 1 & 1& 1 & \textbf{3} & 37.22 & 1.24 & 1.56 & 6.50 \\ 
\hline\hline
Human pair & Fig.~\ref{fig:eg_faust_texture} & 5K & 2 & - & - & \textbf{4} & 286.2  & - & - & 7.21 \\
Tiger pair & Fig.~\ref{fig:randIni_seg}  & 5K & 1 & - & - & \textbf{2} & 311 & - & - & 6.27 \\
Gorilla pair & Fig.~\ref{fig:res:gorilla} & 5K & 2 & - & - & \textbf{4} & 173.4 & - & - & 9.96 \\
Cat pair & Fig.~\ref{fig:eg_cat} & 5K & 1 & - & - & \textbf{2} & 156.7 & - & - & 10.8 \\
Centaur pair & Fig.~\ref{fig:eg_centaur} & 5K & \textbf{2} & - & - & \textbf{2} & 140.6 & - & - & 11.5\\
\hline
\end{tabular}
\end{table}

\begin{figure}[!t]
    \centering
    \begin{overpic}[trim=8cm 50cm 0cm 28cm,clip,width=1\linewidth,grid=false]{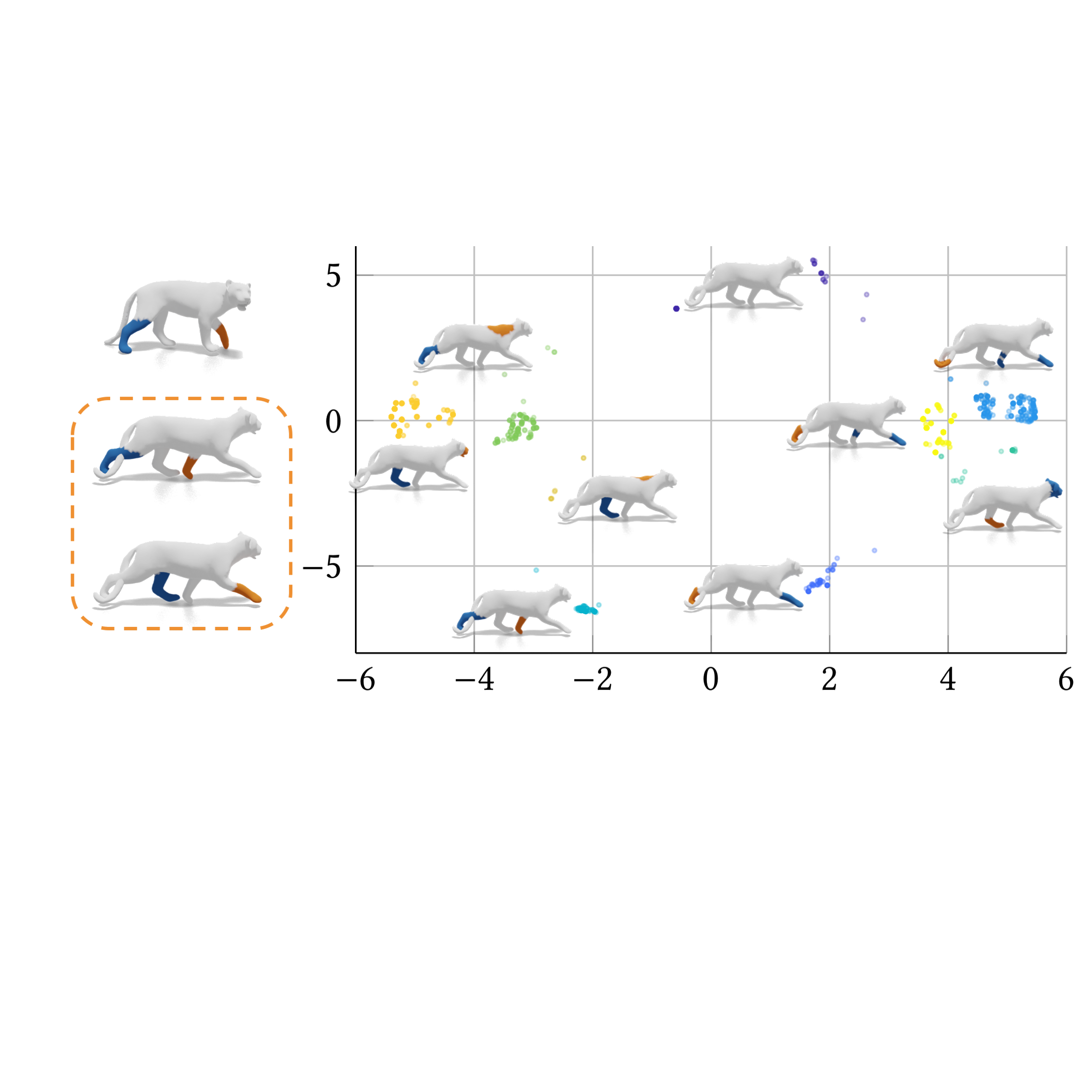}
    \put(8,44){\footnotesize Source}
    \put(8,32){\footnotesize Target}
    \put(-1,5){\footnotesize (Our computed maps)}
    \put(31,46){\footnotesize MDS embedding of 1000 random maps after refinement}
    \end{overpic}\vspace{-9pt}
    \caption{\emph{Left:} For this pair of tigers, we apply our method and obtain the direct and the symmetric map visualized via segment transfer. \emph{Right:} We randomly generate 1000 point-wise maps and apply {\zo} for refinement. We then embed the refined maps in 2D using MDS. We apply a simple K-means to cluster the 2D points and color each point/map accordingly. For each cluster, we show one representative map that is closest to the direct/symmetric map. See Fig.~\ref{fig:randIni_map} in Appendix for the corresponding map visualization via color transfer.}
\label{fig:randIni_seg}
\end{figure}

\begin{figure}[!t]
\centering
\begin{overpic}[trim=1cm 0cm 0cm 0cm,clip,width=1\linewidth,grid=false]{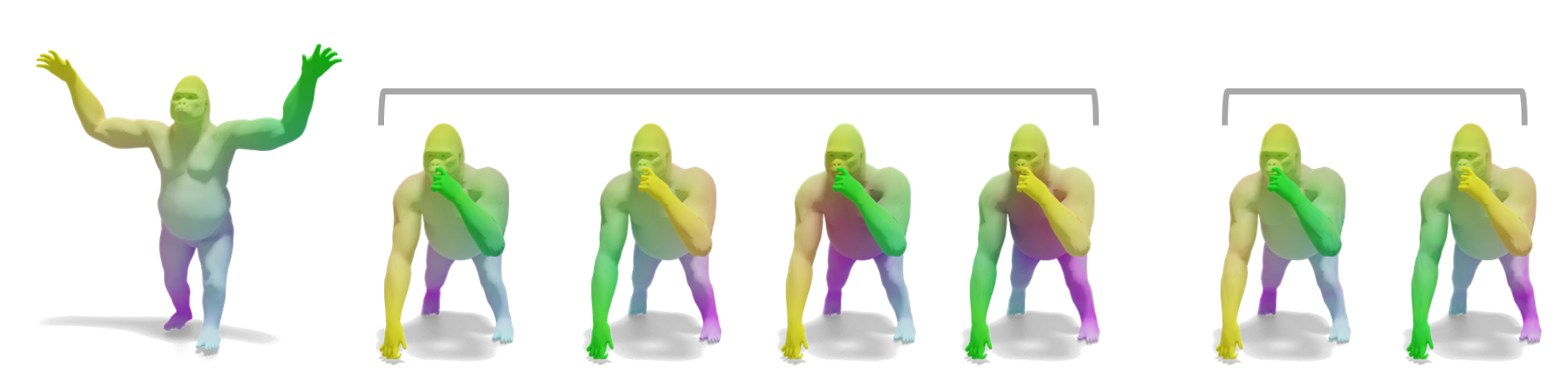}
\put(8,22){\footnotesize Source}
\put(35,22){\footnotesize MapTree (\textbf{ours})}
\put(75,22){\footnotesize randIni + {\zo}}
\end{overpic}\vspace{-12pt}
\caption{For this pair of gorillas, our method obtains four semantically meaningful maps. As a comparison, applying {\zo} to 500 random point-wise maps obtains two maps. }\label{fig:res:gorilla}
\end{figure}

For our quantitative evaluation, we would like to test how many distinct high-quality near isometric maps can be recovered by different methods. To simplify this test, we only count near isometric maps associated with a symmetry and do not consider other low energy maps such as twisted maps as solution. Since we only have ground truth data for some of the maps, we use visual inspection to decide if the maps are smooth and correspond to a symmetry and the geodesic distortion metric to confirm that they are low energy.
We compiled a small test dataset consisting of 7 interesting shapes for the recovery of maps from the shape to itself and 5 shape pairs for the recovery of maps between two shapes.
In this test, we compare to three competitors: 1) our baseline consisting of 500 random initializations + {\zo} 2) IntSymm~\cite{Nagar_2018_ECCV} and 3) OrientRev~\cite{ren2018continuous} + {\zo}. The last two competitors are listed for completeness only, as they are the current state of the art methods for single-solution symmetry detection. These methods can only recover a maximum of one map by design.
Our results are shown in Table~\ref{tb:res:selfSymm}.
We can observe that our method is the only algorithm that can reliably recover multiple high-quality near isometric maps.
The result of random initialization + {\zo} recovers fewer maps except for the cup and the centaur pair.

We provide an example visualization to show why random initialization + {\zo} has difficulty recovering multiple distinct maps using multi-dimensional scaling (MDS) to embed the refined maps in 2D (see Fig.~\ref{fig:randIni_seg}). We cluster the embedded points/maps and show the best map of each cluster. To better visualize the map, we show how each map transfers the front-left leg and the back-right leg. The visualization of the complete map via color transfer can be found in Fig.~\ref{fig:randIni_map} in the appendix. We can notice that most of the maps are of low-quality, e.g., they map a leg to the back or to the head. Some maps are also not smooth and map one leg to two legs. Only one cluster contains high quality maps, but none of the maps recover the symmetry. This shows that the size of the map search space is too large so that it is almost impossible to sample the random initial maps that can lead to different high-quality meaningful maps. In contrast, our method recovers the two meaningful (direct and symmetric) maps as shown in the insets of Fig.~\ref{fig:randIni_seg} and Fig.~\ref{fig:randIni_map}. We show another example where our tree structure can help to find more meaningful maps on a pair of gorillas in Fig.~\ref{fig:res:gorilla}.

\begin{figure}[!t]  
    \centering
    \begin{overpic}[trim=5cm 4.5cm 7cm -3.5cm,clip,width=1\linewidth,grid=false]{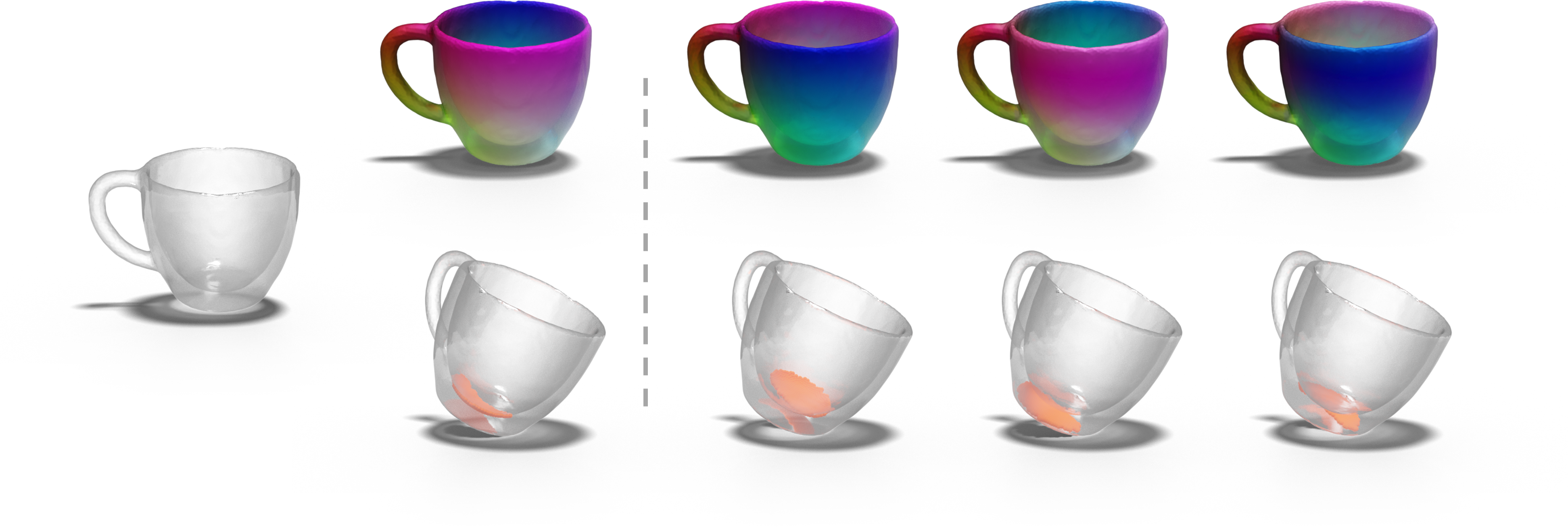}
    \put(-1,25){\footnotesize Double-wall glass}
    \put(26,36){\footnotesize Source}
    \put(26,17){\footnotesize Samples}
    \put(45,37.5){\footnotesize Left-to-right}
    \put(45,34.5){\footnotesize (w.r.t handle)}
    \put(66,36){\footnotesize Inside-out}
    \put(84,37.5){\footnotesize Left-to-right +}
    \put(85.5,34.5){\footnotesize inside-out}
    \end{overpic}
    \vspace{-18pt}
    \caption{Symmetry of a double-wall glass cup computed with our method. \emph{Top}: three self-symmetric maps visualized via color transfer. \emph{Bottom}: to better visualize the inside-out symmetries, a small set of samples on the bottom of the \emph{inner wall} of the cup are highlighted in orange. We then visualize how these samples are mapped according to the symmetries shown in the top row. Note that for the two inside-out symmetric maps, the mapped samples are on the bottom of the outer wall.}
    \label{fig:mtd:eg_cup}
\end{figure}

\begin{figure}[!t]  
    \centering
    \begin{overpic}[trim=0cm 0cm 0cm -4cm,clip,width=1\linewidth,grid=false]{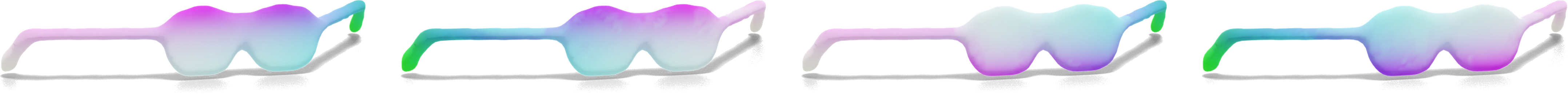}
    \put(10,9){\footnotesize Source}
    \put(32,9){\footnotesize Left-to-right}
    \put(55,9){\footnotesize Upside-down}
    \put(80,11){\footnotesize Left-to-right + }
    \put(80,8){\footnotesize Upside-down}
    \end{overpic}
    \vspace{-18pt}
    \caption{Symmetries detected with our method on a ``glasses'' shape visualized via color-transfer. Observe the detected left-to-right, upside-down, and the simultaneous double symmetry flip.}
    \label{fig:mtd:eg_glasses}
\end{figure}

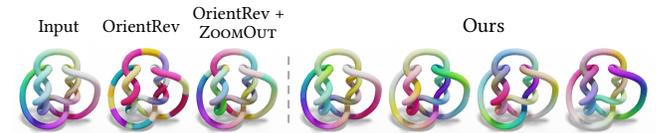
\begin{figure}[!t]
\centering
\begin{overpic}[trim=2cm 1.2cm 2cm -3cm,clip,width=1\linewidth,grid=false]{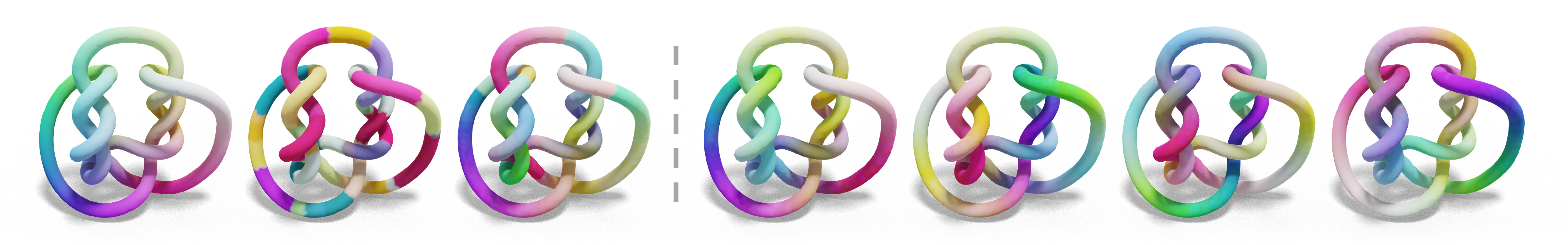}
\put(4,16){\footnotesize Input}
\put(14,16){\footnotesize OrientRev}
\put(28,18){\footnotesize OrientRev +}
\put(29,15){\footnotesize {\zo}}
\put(70,16){\small Ours}
\end{overpic}
\vspace{-18pt}
\caption{For this challenging shape with smooth symmetry group, the IntSymm method fail to output a symmetric map and OrientRev method produce a noisy map with mixed symmetry. As a comparison, our method still outputs several high-quality self-symmetric maps.}\label{fig:res:knot}
\end{figure}

Finally, we showcase three models to visualize interesting maps that can be recovered by our method.
For example, on the cup model (Fig.~\ref{fig:mtd:eg_cup}) we can recover an interesting inside-out map and on the glasses (Fig.~\ref{fig:mtd:eg_glasses}) there are left-to-right and upside-down symmetries. 
Fig.~\ref{fig:res:knot} shows a challenging case of a knot shape with a smooth/continuous symmetry group. In this case, IntSymm even fails to output a single map (instead, it outputs an identity map) and ``OrientRev'' produces a rather noisy map, while our method is still able to output several accurate self-symmetric maps.



\subsection{Non-rigid shape matching\label{sec:res:shape_matching}}
We apply our map tree exploration Algorithm~\ref{alg:mtd:mapTree} to the SHREC'19 Challenge dataset \cite{SHREC19}, which involves 430 shape pairs from 44 shapes with different mesh resolution, triangulation, and partiality. We consider this to be one of the most challenging datasets available for shape matching and we use it to demonstrate that our MapTree algorithm can significantly outperform the state-of-the-art methods in traditional single solution shape matching.
We compare to the best and most closely-related shape matching methods and list the average error, geodesic distortion, and runtime of each method in Table~\ref{tb:res:shrec19_maptree}.

\setlength{\tabcolsep}{0.9em}
\begin{table}[!t]
\caption{
\revised{
Quantitative results on the SHREC'19 Challenge. For the map-tree based methods, the geodesic distortion (shown in brackets) is averaged over \emph{all produced maps}.}
}\vspace{-6pt}
\label{tb:res:shrec19_maptree}
\footnotesize
\begin{tabular}{|l|c|c|c|}
\hline
\multicolumn{1}{|c|}{Method\,\textbackslash\, Measurement} & \begin{tabular}[c]{@{}c@{}}Error\\ $(\times 10^{-3})$\end{tabular} & \begin{tabular}[c]{@{}c@{}}Geodesic\\ Distortion\end{tabular} & \begin{tabular}[c]{@{}c@{}}Runtime\\ (sec)\end{tabular} \\ \hline\hline
BIM~\textcolor{gray}{\cite{kim2011blended}} & 81.2 & 2.314 & 169 \\
\revised{Genetic~\textcolor{gray}{\cite{sahilliouglu2018genetic}}} & \revised{339.4} & \revised{5.458} & \revised{25.11}\\
OrientOp (Ini)~\textcolor{gray}{\cite{ren2018continuous}} & 141.2 & 1.953 &  5.81\\
Ini + ICP~\textcolor{gray}{\cite{ovsjanikov2012functional}}   & 130.2 & 1.596  & 3.82\\
Ini + PMF~\textcolor{gray}{\cite{vestner2017product}} & 143.5 & 1.862 &  429\\
Ini + BCICP~\textcolor{gray}{\cite{ren2018continuous}} & 85.0 & 1.147 & 209 \\
Ini + RHM~\textcolor{gray}{\cite{ezuz2018reversible}} & 125.2 & 1.403 &  25.2 \\
Ini + {\zo}~\textcolor{gray}{\cite{zoomout}} & 118.5 & 1.367 & 0.38 \\\hline\hline
\textbf{MapTree (Ours)} - GT selection & \textbf{28.42} & \textbf{0.93 }(1.16)  & 35.5\\ 
\textbf{MapTree (Ours)} - Auto selection & \textbf{39.99} & \textbf{0.94} (1.16) & 35.5\\ \hline
\end{tabular}
\end{table}

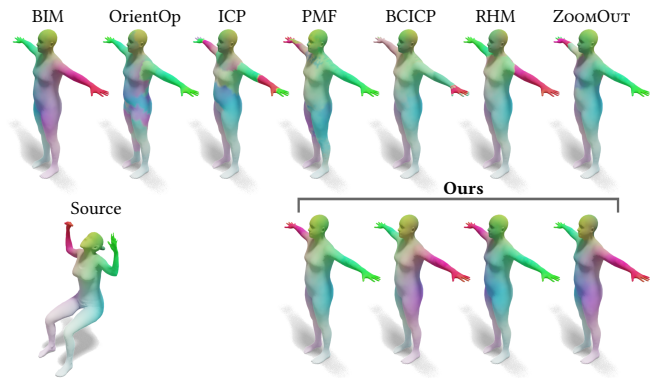
\begin{figure}[!t] 
    \centering
    \begin{overpic}[trim=0cm 0cm 0cm -2cm,clip,width=1\linewidth,grid=false]{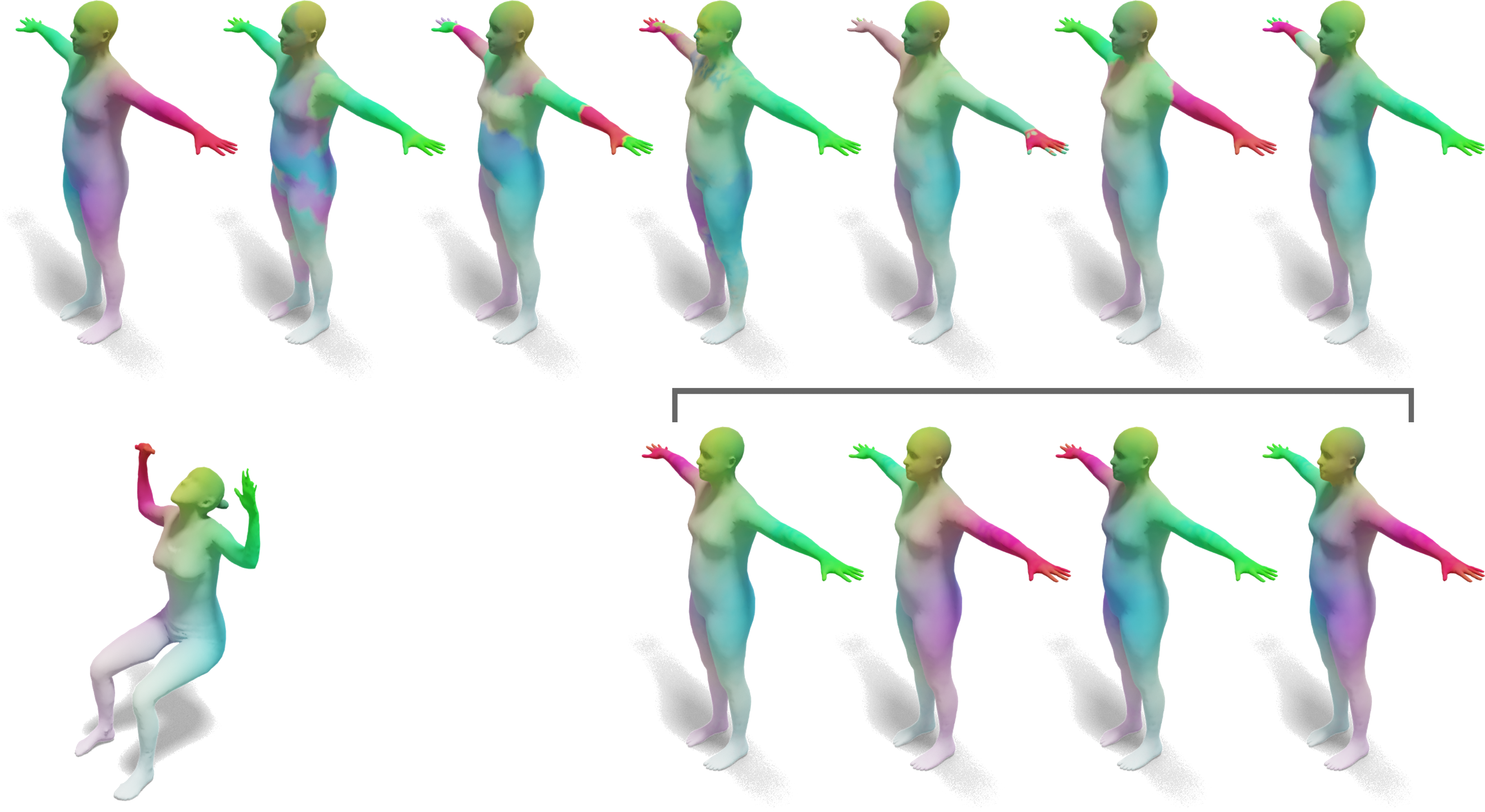}
    \put(10,26){\footnotesize Source}
    \put(68,29.2){\footnotesize \textbf{Ours}}
    \put(4,56){\footnotesize BIM}
    \put(16,56){\footnotesize OrientOp}
    \put(33,56){\footnotesize ICP}
    \put(46,56){\footnotesize PMF}
    \put(59,56){\footnotesize BCICP}
    \put(73,56){\footnotesize RHM}
    \put(85,56){\footnotesize {\zo}}
    \end{overpic}
    \vspace{-23pt}
    \caption{Qualitative comparison of maps obtained with different methods. When the initialization (OrientOp) is of poor quality, the refinement methods including ICP, PMF, BCICP, RHM, and {\zo} fail to recover a good map. Meanwhile, BIM outputs a back-to-front symmetric map which is not preferable for symmetry detection.
    In contrast, our method can output a set of smooth maps with different orientations for different applications.}
    \label{fig:res:shrec19_eg}
\end{figure}

In single-solution shape matching the algorithms are expected to output a single map with the direct orientation (i.e, map the left of a shape to the left, assuming there is a left-to-right symmetry). Competing methods are: BIM~\cite{kim2011blended}, and several functional map based methods, where we use the best known automatic initialization~\cite{ren2018continuous}, and refine it using different methods, such as ICP~\cite{ovsjanikov2012functional}, PMF~\cite{vestner2017product}, BCICP~\cite{ren2018continuous}, RHM~\cite{ezuz2018reversible}, and {\zo}~\cite{zoomout}. Fig.~\ref{fig:res:shrec19_eg} shows an example of the maps obtained from different methods.
\revised{In Table~\ref{tb:res:shrec19_maptree}, we also compare to the genetic algorithm~\cite{sahilliouglu2018genetic}, where sparse correspondences are computed and interpolated. 
As shown in the table, the average direct error (0.339) and the average error that allows symmetry flips (0.289) are significantly higher than ours (0.047). In Fig.~\ref{fig:eg_genetic} we visualize the computed maps by the genetic algorithm on three example shape pairs. Note that the genetic algorithm can produce accurate sparse correspondences but fails to produce high-quality and smooth dense maps.}

We propose two versions of our algorithm: semi-automatic and automatic. In the semi-automatic method we assume that the user selects the best map among the available maps. We simulate this user selection by selecting the left-to-left direct map closest to the ground truth and denote this method as \emph{GT selection}. The average error of our method is 28 while the best baseline gives 81.

The main idea of our automatic selection algorithm is to use cycle consistency. We propose a simple approach to achieve this in the spirit of~\cite{huang2014functional}, but using the fact that our method produces multiple maps for each shape pair.  Specifically, for every pair $S_1$, $S_2$ in the benchmark, our algorithm produces 10 maps $\{ C_{12}^{i} \}_{i = 1}^{10} $. We perform an automatic map selection in two steps.
We first initialize the selection for each shape pair with the map that minimizes the orientation-preservation error introduced in \cite{ren2018continuous}. We then iteratively update this selection by considering triplets of shapes, and by minimizing the following cycle consistency energy:
%
%
\begin{equation}\label{eq:err_cicleconsistency}
\begin{split}
E_{\text{CyCons}}\big(C_{12}^{i}\big) =\; & \sum_{j \in J} \Vert C_{1j}^{\star}C_{j2}^{\star} - C_{12}^{i} \Vert_F + \Vert C_{12}^{i}C_{2j}^{\star} - C_{1j}^{\star} \Vert_F\\ 
+\; & \Vert C_{j1}^{\star}C_{12}^{i} - C_{2j}^{\star} \Vert_F
\end{split}
\end{equation}
where $J$ is the set of all shape pairs in the benchmark that form a triplet with 
$S_1$ and $S_2$. Thus, we update the map selection for the pair $S_1,S_2$ by selecting $C_{12}$ among the 10 possible maps, the one that minimizes this energy, while keeping the selection for the other shape pairs (identified by the superscript $\star$) in Equation~\eqref{eq:err_cicleconsistency} fixed from the previous iteration. This very simple strategy converges remarkably fast, within 2-3 iterations over all map pairs.

The results of the cycle consistency-based automatic selection are also shown in Table~\ref{tb:res:shrec19_maptree} as ``MapTree (ours) - Auto selection.'' This combination makes our method a fully automatic method for shape matching, which achieves the state-of-the-art accuracy with a 50\% improvement over the best existing baseline.
It is also interesting to note that the average geodesic distortion of all the produced maps across the complete dataset is 1.16. This is remarkable, because it shows that most of our recovered maps have smaller geodesic distortion than the direct map recovered by competing methods. It is also an indication that geodesic distortion alone is not sufficient to distinguish between the direct map and other high quality maps.


\setlength{\tabcolsep}{0.55em}
\begin{table}[!t]
\caption{\revised{
Self-symmetry detection on SHREC'19. We compare the accuracy and different smoothness measurement (including geodesic distortion, Dirichlet energy, and conformal distortion) of the self-symmetric maps from different baselines.}}
\label{tb:shrec19:selfsymm}\vspace{-6pt}
\footnotesize
\begin{tabular}{|l|c|c|c|c|c|}
\hline
\multicolumn{1}{|c|}{\begin{tabular}[c]{@{}c@{}}Methods \textbackslash\, \\ Measurement\end{tabular}} &
  \multicolumn{1}{c|}{\begin{tabular}[c]{@{}c@{}}Accuracy\\ ($\times 10^{-3})$\end{tabular}} &
  \multicolumn{1}{c|}{\begin{tabular}[c]{@{}c@{}}GeoDist\\ ($\times 10^{2})$\end{tabular}} &
  \multicolumn{1}{c|}{\begin{tabular}[c]{@{}c@{}}\revised{Dirichlet}\\ \revised{Energy}\end{tabular}} &
  \multicolumn{1}{c|}{\begin{tabular}[c]{@{}c@{}}\revised{Conformal}\\
  \revised{Distortion}\end{tabular}} &
  \multicolumn{1}{c|}{\begin{tabular}[c]{@{}c@{}}Runtime\\ (sec)\end{tabular}} \\ \hline\hline
BIM  & 83.69 & \textbf{1.418} &  \textbf{3.278} & 1.970 & 164  \\
GroupRep  & 311.1  & 5.254 & 13.41 & 7.787 & 3.95 \\
IntSymm & \textbf{62.50} & 1.945 & 12.17 & 7.123 & 2.05 \\ 
OrientRev (Ini) & 137.2  & 4.682  & 22.07 & 12.69 & 0.52 \\
Ini + ICP  & 108.9  & 3.604 & 10.49 & 6.235  & 8.33 \\
Ini + PMF  & 119.4 & 2.444 & 15.98 & 9.605 & 425  \\
Ini + RHM  & 118.7 & 4.166 & 7.352 & 4.369 & 28.7 \\
Ini + BCICP & 96.72 & 2.466 & 5.633 & 3.741 & 157  \\
Ini + ZoomOut & 80.30 & 2.858 & 6.601 & 3.838 & 6.58 \\ \hline\hline
\textbf{MapTree} - GT & \textbf{39.62} & 1.512 & 3.763 & \textbf{0.949}  & 65.2 \\
\textbf{MapTree} - Auto& \textbf{47.48} & 1.507 & 3.833 &  \textbf{0.929} & 65.2 \\ \hline
\end{tabular}
\end{table}

We perform a second large scale experiment on more than 200 pairs of SCAPE human shapes, where we also achieved a significant improvement over the best baseline by 23\%. The details of the test on SCAPE and some additional visualizations including the geodesic distortion and texture transfer can be found in Appendix~\ref{append:add_res}.

\subsection{Self-symmetry detection.}

Our map tree algorithm can also be used to improve upon the state-of-the-art in single-solution self-symmetry detection. Table~\ref{tb:shrec19:selfsymm} reports a quantitative evaluation on self-symmetry detection on the SHREC'19 dataset that consists of 44 shapes with different poses. 
Our method produces eight self-maps per shape. We then select the preferable self-symmetric map among the 8 maps in two ways: (1) "GT selection": we use the given ground-truth self-symmetric map to select the map with the smallest distance to the ground-truth. (2) "Auto selection": we simply select the map that is far from the identity map with a small Laplacian-commutativity error. Note that this is a fully automatic method. Comparing to the best baseline IntSymm~\cite{Nagar_2018_ECCV} on this dataset, our method achieves 24\% improvement on accuracy with better geodesic distortion. 
\revised{Also note that, the maps produced by our method achieve the smallest conformal distortion compared to all the baselines, and achieve comparable Dirichlet energy and geodesic distortion to BIM, while outperforming all other methods.}

\begin{figure}[!t] 
    \centering
    \begin{overpic}[trim=3cm 0cm 2cm -2cm,clip,width=1\linewidth,grid=false]{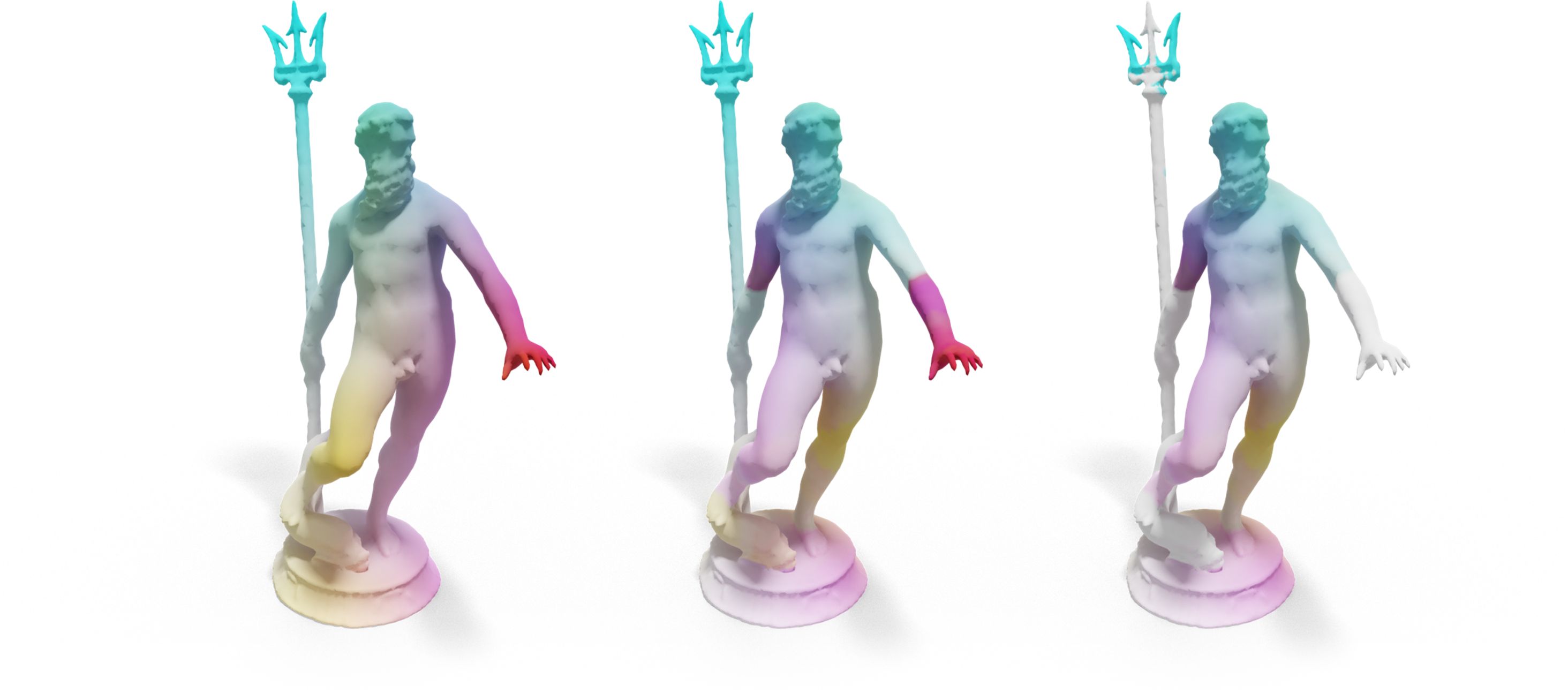}
    \put(15,49){\small Input}
    \put(37,49){\small Obtained self-map}
    \put(68,49){\small Partial symmetry}
    \end{overpic}
    \vspace{-1cm}
    \caption{Partial intrinsic symmetries of the Neptune statue. \emph{Left}: the input model is not strictly symmetric due to the topology change. \emph{Middle}: we run our algorithm directly and obtain the self-map visualized via color transfer. \emph{Right}: we only highlight the shape regions that have intrinsic symmetry.}
    \label{fig:res:eg_neptune}
\end{figure}

\paragraph{Partial intrinsic symmetries}
We also applied our self-symmetry exploration algorithm to shapes with \emph{partial} intrinsic symmetries without modification. Fig.~\ref{fig:res:eg_neptune} shows such an example, where the input Neptune model is not strictly symmetric. We use our method and obtain the self-map as visualized in the middle of Fig.~\ref{fig:res:eg_neptune}. We can see that most of the body region contains a symmetry, while the remaining part does not due to the topology change of having the hand attached to the spear and the feet attached to the statue's base. We can then extract the region with symmetry from the map, as shown in the right of Fig.~\ref{fig:res:eg_neptune}, where only the region that contains symmetry is colored w.r.t. the symmetric vertex via color transfer, and the remaining region is colored white. See Fig.~\ref{fig:res:eg_ballet},~\ref{fig:res:eg_memento} and Fig.~\ref{fig:res:eg_angle} in Appendix for more examples. We observe that our method produces high quality results in these challenging cases, due to the robustness and strong convergence power of our map refinement even if, e.g., the bijectivity constraint is not strictly satisfied.

\subsection{Robustness}
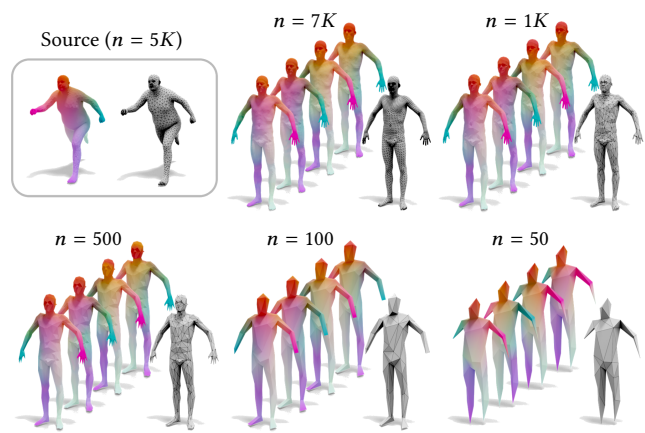
\begin{figure}[!t] 
    \centering
    \begin{overpic}[trim=1cm 0cm 0cm 0cm,clip,width=1\linewidth,grid=false]{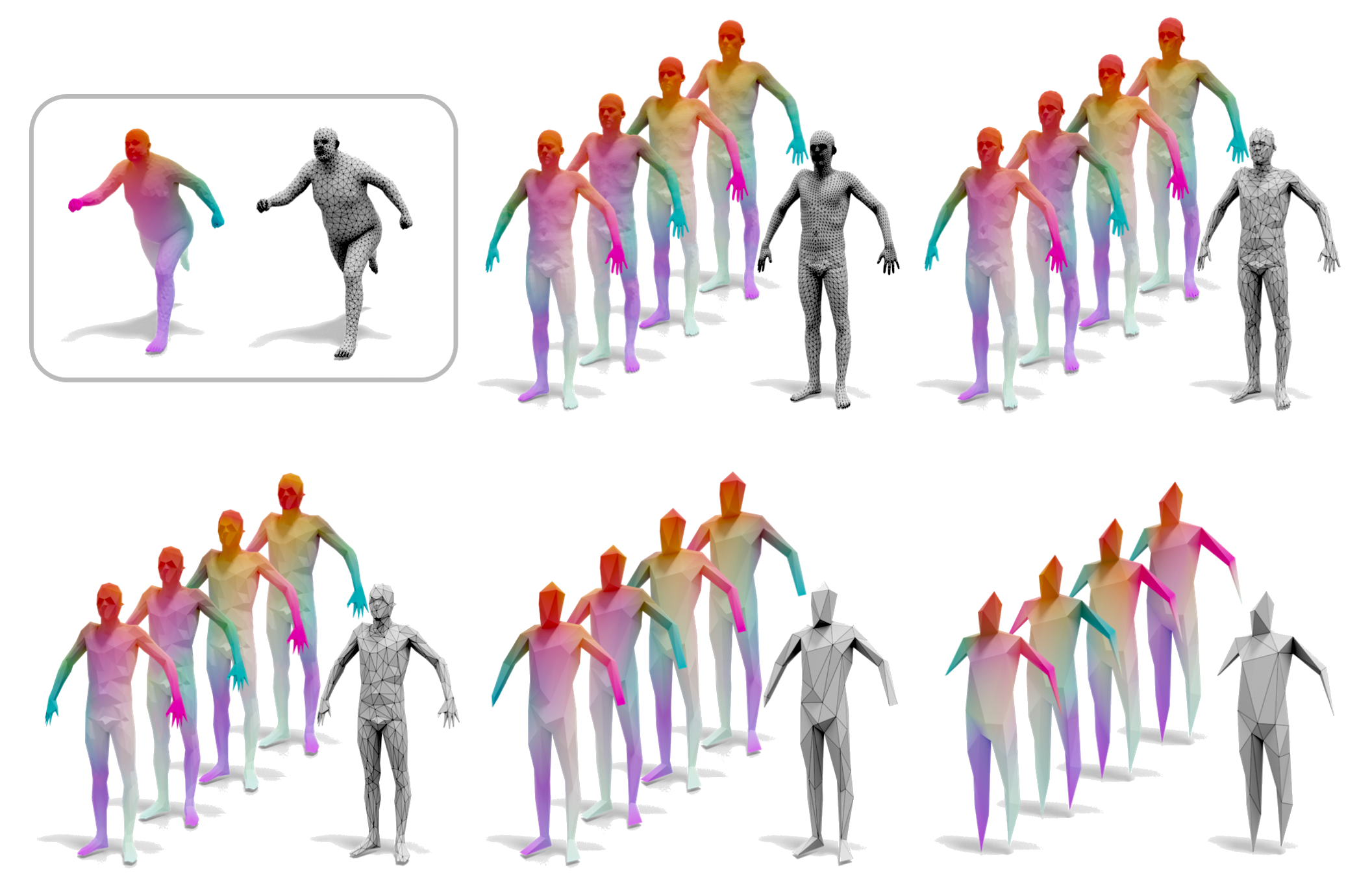}
    \put(6,62){\small Source ($n = 5K$)}
    \put(42,65){\small $n = 7K$}
    \put(75,65){\small $n = 1K$}
    \put(8,31){\small $n = 500$}
    \put(41,31){\small $n = 100$}
    \put(76,31){\small $n = 50$}
    \end{overpic}
    \vspace{-0.8cm}
    \caption{\revised{
    Robustness w.r.t. decimation. For a human shape with different triangulation (source from SHREC'19, target from FAUST), we downsample the target shape to the number of $n$ vertices. We use the \emph{same} set of parameters and apply our MapTree algorithm. Note that even at the resolution of $n=100$, our method still obtains four reasonable maps that are consistent with other resolutions.
    }}
    \label{fig:res:eg_decimation}
\end{figure}

\revised{ Our method is robust w.r.t. mesh triangulation and decimation. Specifically, the SHREC'19 dataset consists of 44 shapes with different mesh resolution and triangulation. Table~\ref{tb:res:shrec19_maptree} and~\ref{tb:shrec19:selfsymm} shows that our method is robust w.r.t. mesh triangulation. We further test the robustness of our method w.r.t. decimation (see Fig.~\ref{fig:res:eg_decimation}). For a given shape pair, we downsample the target shape to different number of vertices. Our method still produces accurate and consistent maps.}

\subsection{Parameters}
For our MapTree construction algorithm, we set $\epsilon$ in Algorithm~\ref{alg:mtd:enumeration} to 1, and the $\epsilon_1$ and $\epsilon_2$ in the pruning rules to 0.5 for most of our tests. For some cases we also set $\epsilon_1, \epsilon_2$ to larger values to allow more leaves to be explored with a prior knowledge that the investigated shape (pair) contains more complicated symmetries. We stop the tree construction process when the leaf functional map has size larger than 10 (for non-human shapes) or 20 (for human shapes).

\subsection{Runtime \& Implementation}
We implemented the proposed methods in MATLAB, and conducted the experiments on a workstation with a 3.10GHz processor and 64 GB memory. The (average) runtime for different tests in seconds is reported (see Table~\ref{tb:res:selfSymm}, \ref{tb:res:shrec19_maptree}, \ref{tb:shrec19:selfsymm}). We used the sub-sampling strategy to speed-up our MapTree algorithm similar to the one used in~\cite{zoomout}. Specifically, we only sample a small number of points (e.g., 200-500) on the shapes to compute a functional map in our MapTree algorithm. In the vast majority of cases our complete pipeline takes 60 seconds end-to-end.  We will release our complete implementation for full reproducibility of all of the results.

\section{Conclusion, Limitations \& Future Work }
In this paper, we proposed a novel approach to recover multiple high quality correspondences between a shape or a shape pair. To achieve this,
we propose a compact tree structure to efficiently organize and explore map space without requiring any landmarks or initialization. We presented a large variety of experiments demonstrating that our method can be used to explore the self-symmetry space or even the partial intrinsic symmetry of a single shape, and to explore the map space between a shape pair, which is not possible for the previous methods. We also show the quantitative results on challenging datasets showing that our method can be used as an automatic shape matching technique that does not require any descriptors or landmarks. The results show that our method significantly outperforms current state-of-the-art. Our algorithm is easy to implement with a small set of parameters. 

Our method still has some limitations. First, we only target discrete sets of maps and symmetries, and do not consider continuous spaces such as the symmetry group of a sphere. Second, while outputting a set of high-quality maps can be beneficial, for the specific application of shape matching, it is not trivial to correctly select the preferred map, even from a small set. 
\revised{Third, we only consider manifold meshes in our setting. However, using a Laplace-Beltrami discretization such as~\cite{Belkin09} or a recent work{~\cite{sharp2020Laplacian}}, our method could be directly generalized to non-manifold meshes or even point clouds. For partial matching, ~\cite{rodola2017partial} can be helpful to design partial functional map expansion rules to potentially extend our work.} 

\revised{We also plan to investigate potential applications that can exploit our map tree structure. For example, Fig.~\ref{fig:res:garment} in the Appendix shows that the multiple detected maps can help the shape modeling process. Meanwhile, Fig.~\ref{fig:mtd:eg_table} shows that the global symmetry information is encoded in the relatively low frequencies. This suggests that our map tree structure encodes level-of-symmetry which might be useful for shape modeling and, e.g., making the triangulation of a given shape symmetric w.r.t. multiple symmetry axes.}

\begin{acks}
The authors would like to thank the anonymous reviewers for their valuable comments and helpful suggestions.
Parts of this work were supported by the KAUST OSR Award No. CRG-2017-3426, and the ERC Starting Grant No. 758800 (EXPROTEA).
\end{acks}

\bibliographystyle{ACM-Reference-Format}
\bibliography{bibliography}

\appendix
\input{proofs.tex}

\input{additional_results.tex}
\end{document}

%% file: header.tex
\usepackage{booktabs} 

\usepackage[ruled]{algorithm2e} 

\SetAlFnt{\small}
\SetAlCapFnt{\small}
\SetAlCapNameFnt{\small}
\SetAlCapHSkip{0pt}

\graphicspath{{figures/}}
\usepackage{bbold} 
\usepackage{amsfonts} 
\usepackage{amsmath} 
\usepackage{amssymb}  
\usepackage{overpic}
\usepackage{pgfplots}
\usepackage{pgf,tikz}
\usepackage{bm}
\usepackage{wrapfig}
\usepackage{multirow}
\usepackage{mathtools}
\usepackage{enumerate}

\usepackage{verbatim}

\SetAlFnt{\small}
\SetAlCapFnt{\small}
\SetAlCapNameFnt{\small}
\SetAlCapHSkip{0pt}

\newcommand{\M}{\mathcal{M}}

\newcommand{\N}{\mathcal{N}}

\newcommand{\LM}{L^2(\M)}
\newcommand{\LN}{L^2(\N)}
\renewcommand{\C}{\mathbf{C}}

\renewcommand{\phi}{\varphi}

\DeclareMathOperator*{\argmax}{arg\,max}

\usepackage{soul}

\usepackage{listings}
\usepackage{color} 
\definecolor{mygreen}{RGB}{28,172,0} 
\definecolor{mylilas}{RGB}{170,55,241}

\newtheorem{theorem}{Theorem}[section]

\usepackage{amssymb}

\newcommand*\circled[1]{\tikz[baseline=(char.base)]{
            \node[shape=circle,draw,inner sep=1pt] (char) {#1};}}
\usepackage[shortlabels]{enumitem}
\usepackage{kantlipsum}

%% file: figures/eg_table.tex
    \begin{overpic}[trim=5cm 2cm 8cm 0cm,clip,width=1\linewidth,grid=false]{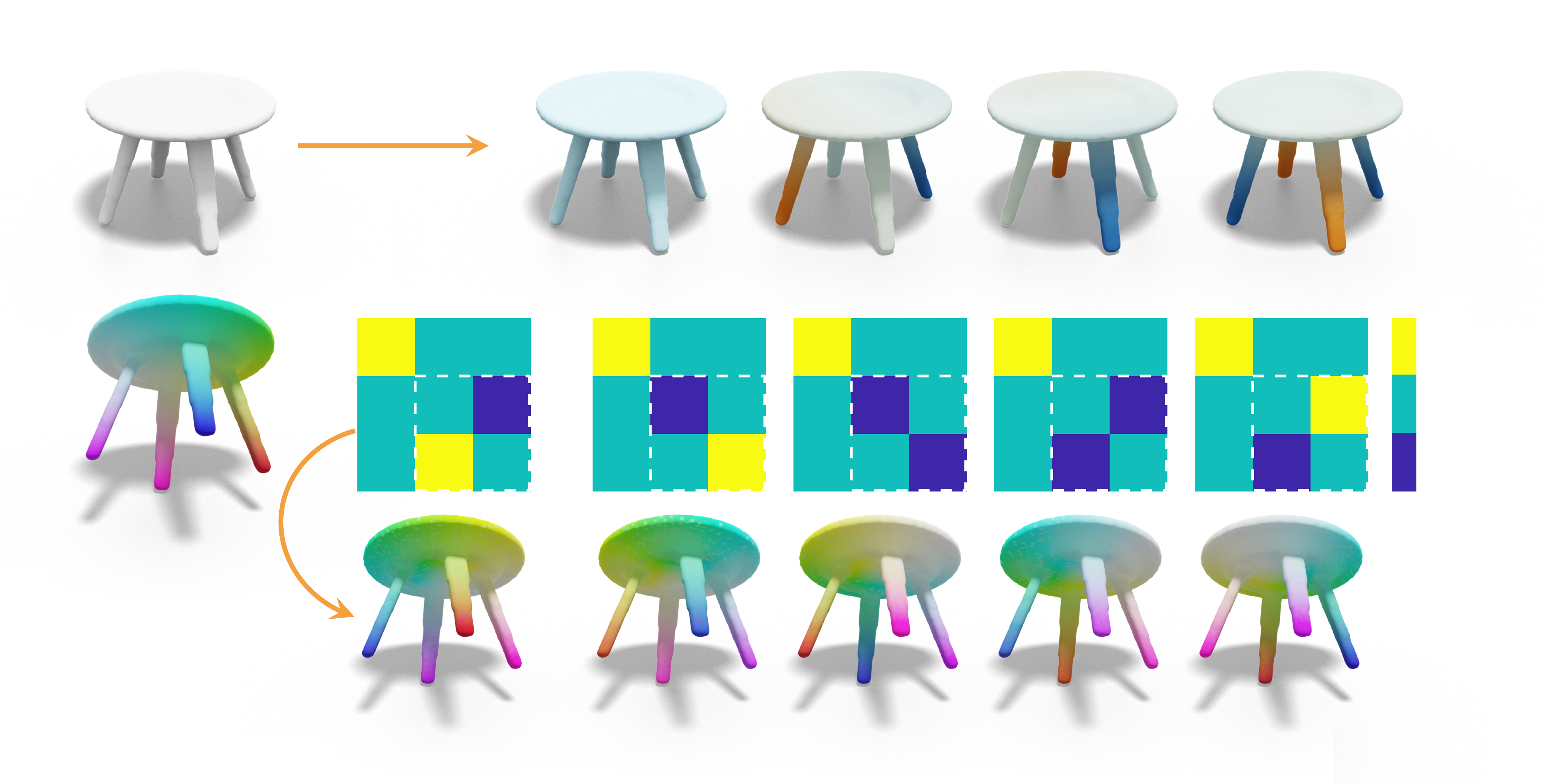}

    \put(4,52){\footnotesize Input $S$}
    \put(15,46){\footnotesize $\Delta_S \phi_k = \lambda_k \phi_k$}
    \put(39, 54){\footnotesize $\phi_1$}
    \put(55, 54){\footnotesize $\phi_2$}
    \put(71, 54){\footnotesize $\phi_3$}
    \put(89, 54){\footnotesize $\phi_4$}
    
    \put(36, 51){\footnotesize $\lambda_1=0$}
    \put(50, 51){\footnotesize $\lambda_2=13.01$}
    \put(67, 51){\footnotesize $\lambda_3=13.04$}
    \put(86, 51){\footnotesize $\lambda_4=15$}
    
    \put(22,33){\footnotesize $\phi_1$}
    \put(26,33){\footnotesize $\phi_2$}
    \put(30,33){\footnotesize $\phi_3$}
    
    \put(34,30){\footnotesize $\phi_1$}
    \put(34,25){\footnotesize $\phi_2$}
    \put(34,20){\footnotesize $\phi_3$}

    \put(0,34){\footnotesize Source}
    \put(0,12){\footnotesize Convert to a}
    \put(0,9){\footnotesize pointwise map}
    
    \put(100,30){\footnotesize 1}
    \put(100,25){\footnotesize 0}
    \put(99,20){\footnotesize -1}
    \end{overpic}

%% file: proofs.tex
\newtheorem*{theorem*}{Theorem}
\section{Theoretical analysis}\label{append:proof}
\subsection{Proof of Theorem~\ref{thm:bound}}
Suppose we are given two compact surfaces $M, N$ and two smooth maps $T_1, T_2: N \rightarrow M$. We consider the pullback operators $T_{F1}, T_{F2}: \mathcal{L}_2(M) \rightarrow \mathcal{L}_2(N)$, defined as $T_{F1}(f) = g_1$, where $g_1(x) = f(T_1(x))$, and similarly $T_{F1}(f) = g_2$, where $g_2(x) = f(T_2(x)).$

We will call $T_1$ and $T_2$ $\delta$-close if the following is satisfied: for each $x\in N$, $d^{M} (T_1(x), T_2(x)) < \delta$ for some $\delta \ge 0$. Here $d^{M}$ denotes the geodesic distance on $M$. 

Now let $\phi_i$ be an eigenfunction of the Laplace-Beltrami operator corresponding to eigenvalue $\lambda_i \ge 0$. Our goal is to relate $T_{F1}(\phi_i)$ to $T_{F2}(\phi_i)$ using the bound above.

Observe that if $g = T_{F1}(\phi_i) - T_{F2}(\phi_i)$ then 
\begin{align}
 g(x) = \phi_i\big(T_1(x)\big) - \phi_i\big(T_2(x)\big)
\end{align}

Using the results from \cite{donnelly2006eigenfunctions,arnaudon2017gradient} we recover the following bound: whenever $M$ is a smooth compact manifold, there exists a universal constant $C$ (depending on the manifold), such that for any eigenfunction $\phi_i$, its gradient satisfies the following:
\begin{align}
    \|\nabla \phi_i\|_{\infty} \le C \lambda_i
\end{align}

Combining this bound with the assumption that $T_1$ and $T_2$ are $\delta$-close, and using the gradient theorem for the line integral on the surface we get that: 

\begin{align}
    \big\vert \phi_i\big(T_1(x)\big) - \phi_i\big(T_2(x)\big)\big\vert \le \delta C \lambda_i ~\forall~ x\in N.
\end{align}

This implies that:
\begin{align}
    \|T_{F1}(\phi_i) - T_{F2}(\phi_i)\|_2^2 =& \int_{x \in N} \Big(\phi_i\big(T_1(x)\big) - \phi_i\big(T_2(x)\big)\Big)^2 d\mu(x)  \\
    \le & (\delta C \lambda_i)^2 \text{Area}(N)
\end{align}

Now recall that the $i^{\text{th}}$ column of the functional map representation of a pointwise map $T$ encodes the coefficients of $TF(\phi^M_i)$ expressed in $\phi^N_i$. Therefore, by Plancherel's theorem we obtain Theorem~\ref{thm:bound}:
\begin{theorem*}
 Given two pointwise maps $T_1, T_2$ that are $\delta$-close, their functional map matrix representation $C_1, C_2$ will satisfy the following bound $\sum_j (C_1(j,i) - C_2(j,i))^2 < (\delta C \lambda_i)^2 \text{Area}(N)$ for any $i$.
\end{theorem*}

Note that this bound implies that the higher frequencies become more and more ``unstable''. Or, alternatively, in order to disambiguate pointwise maps that are very close, one must consider higher and higher frequencies.

\subsection{Lower bound of functional map representation}
\label{appendix:lower_bound}
Suppose we are given two shapes $\M, \N$, and two pointwise maps between them $T_1, T_2: \N \rightarrow \M$. This induces two functional maps $T_{F1}, T_{F2} : \LM \rightarrow \LN$ given via pull-backs. Specifically for any function $f \in \LM$, the corresponding function $g \in \LN$ is given by $T_{F1}(f) = g$, where $g(y) = f(T_1(y))$ for any $y \in \N$ (and similarly for $T_{F2}$ defined via pull-back with respect to $T_2$).

We define the functional map matrices $\C_1, \C_2$ by expressing $T_{F1}$ and $T_{F2}$ in the Laplace-Beltrami basis. Specifically, $\C_1(i,j) = <T_{F1} (\phi^\M_j), \phi^\N_i>_{\N}$, where $\phi^{\M}_i$ is the $i^{\text{th}}$  eigenfunction of the Laplace-Beltrami operator on $\M$, and $< \cdot, \cdot >_\N$ is the standard $L_2$ inner product for functions on $\N$.

Our goal is to bound the difference between $T_1, T_2$ and the difference between the matrices $\C_1, \C_2$. Specifically we would like to show that if $T_1$ and $T_2$ \emph{are sufficiently different} then the submatrices of $\C_1, \C_2$ will be different as well.

For this we consider any region $\mathcal{R} \subseteq \M$ and denote by $\mathcal{P}_1, \mathcal{P}_2$ its pre-images via $T_1$ and $T_2$. Namely, $\mathcal{P}_1 = \{y \in \N, \text{ s. t. } T_1(y) \in \M\}$. Note that if $\iota_{\mathcal{R}}$ is the indicator function of $\mathcal{R}$ and $\iota_{\mathcal{P}_1}$, $\iota_{\mathcal{P}_2}$, are indicator functions of $\mathcal{P}_1, \mathcal{P}_2$, then we have $T_{F1} (\iota_{\mathcal{R}}) = \iota_{\mathcal{P}_1}$ and $T_{F2} (\iota_{R}) = \iota_{\mathcal{P}_2}$.

Now observe that for any region $\mathcal{R}$ we have:
\begin{align*}
 \|T_{F1} (\iota_{\mathcal{R}}) - T_{F2} (\iota_{\mathcal{R}})\|_2^2 &= \int_{\N} (\iota_{\mathcal{P}_1} (y) - \iota_{\mathcal{P}_1}(y))^2 dy  \\
 &= \text{Area}(\mathcal{P}_1 \ominus \mathcal{P}_2),
\end{align*}
where $\ominus$ denotes the symmetric difference of the two sets $\mathcal{P}_1 \ominus \mathcal{P}_2 = \left(\mathcal{P}_1  \setminus \mathcal{P}_2 \right) \cup \left(\mathcal{P}_2  \setminus \mathcal{P}_1 \right)$.

Suppose that the given maps $T_1, T_2$ are \emph{sufficiently different} in the following sense: there exists a region $\mathcal{R}$ s.t. its two pre-images $\mathcal{P}_1, \mathcal{P}_2$ via $T_1$ and $T_2$ are such that $\text{Area}(\mathcal{P}_1 \ominus \mathcal{P}_2) > A$ where $A$ some constant. Moreover, suppose that the projection of $\iota_{\mathcal{R}}$ onto the first $k_{\M}$ eigenfunctions of the LB operator on $\M$ is accurate up to $\varepsilon$ error: $\|\iota_{\mathcal{R}} - \sum_{i=1}^{k_{\M}} a_i \phi_i^\M\|_2^2<\varepsilon \text{Area}(\mathcal{R}), $ where $a_i = <\phi_i^\M, \iota_{\mathcal{R}}>_{\M}$. Similarly, we assume that the projection of $\iota_{\mathcal{P}_1 \ominus \mathcal{P}_2}$ onto the first $k_{\N}$ eigenfunctions of the LB operator on $\N$ is accurate up to $\delta$ error: $\|\iota_{\mathcal{P}_1 \ominus \mathcal{P}_2} - \sum_{i=1}^{k_{\N}} b_i \phi_i^\N\|_2^2<\delta \text{Area}(\mathcal{P}_1 \ominus \mathcal{P}_2), $ where $b_i = <\phi_i^\N, \iota_{\mathcal{P}_1 \ominus \mathcal{P}_2}>_{\N}$. Finally, for simplicity we will also assume that $T_1, T_2$ are both locally area-preserving, although a more refined variant of this bound can also be established under less restrictive assumptions (e.g., only assuming bijectivity and smoothness).

Now let $\tilde{\mathbf{C}_1}$ and $\tilde{\mathbf{C}_2}$ be the $k_{\N} \times k_{\M}$ leading principal submatrices of $\C_1, \C_2$ (i.e., sub-matrices obtained by taking the first $k_{\N}$ rows and first $k_{\M}$ columns of $\C_1, \C_2$). Our goal is to bound the Frobenius norm: $\|\tilde{\mathbf{C}_1} -  \tilde{\mathbf{C}_2}\|_F^2$ as a function of the constants $A, \varepsilon, \delta$ introduced above.

For this remark that:

\begin{equation}\begin{split}
     & \|T_{F1}(\iota_{\mathcal{R}} - \sum_{i=k_{\M} + 1}^{\infty} a_i \phi_i^\M) - T_{F2}(\iota_{\mathcal{R}} - \sum_{i=k_{\M} + 1}^{\infty} a_i \phi_i^\M)\|_2 \\
     &= \|T_{F1}(\iota_{\mathcal{R}}) - T_{F2}(\iota_{\mathcal{R}}) - T_{F1}( \sum_{i=k_{\M} + 1}^{\infty} a_i \phi_i^\M) - T_{F2}(\sum_{i=k_{\M} + 1}^{\infty} a_i \phi_i^\M)\|_2 \\
     &\ge \|\iota_{\mathcal{P}_1 \ominus \mathcal{P}_2}\|_2 - \|T_{F1}( \sum_{i=k_{\M} + 1}^{\infty} a_i \phi_i^\M)\|_2 - 
     \|T_{F2}( \sum_{i=k_{\M} + 1}^{\infty} a_i \phi_i^\M)\|_2.
     \\
     &= \sqrt{\text{Area}(\mathcal{P}_1 \ominus \mathcal{P}_2)} - \|T_{F1}( \sum_{i=k_{\M} + 1}^{\infty} a_i \phi_i^\M)\|_2 - 
     \|T_{F2}( \sum_{i=k_{\M} + 1}^{\infty} a_i \phi_i^\M)\|_2.
\end{split}\end{equation}

Now recall that by assumption of local area preservation we have $\|T_{F1}(f)\|_2 = \|f\|_2$ for any function $f$. Thus, if we let $\mathbf{a}$ be the first $k_{\M}$ coefficients of $\iota_{\mathcal{R}}$ in the LB basis of $\M$ as introduced above, then:

\begin{align}
    \frac{\|(\tilde{\mathbf{C}_1} - \tilde{\mathbf{C}_2}) \mathbf{a} \|}
    {\|\mathbf{a}\|} \ge  \frac{\sqrt{\text{Area}(\mathcal{P}_1 \ominus \mathcal{P}_2)}(1 - \sqrt{\delta})}{\sqrt{\text{Area}(\mathcal{R})}} - 2\sqrt{\varepsilon}.
\end{align}

Finally, the Frobenius norm is bounded by the L2 norm we have:
\begin{align}
    \|(\tilde{\mathbf{C}_1} - \tilde{\mathbf{C}_2}) \|_{F} &\ge  \|(\tilde{\mathbf{C}_1} - \tilde{\mathbf{C}_2}) \|_{2} = \max_{\mathbf{x}} \frac{\|(\tilde{\mathbf{C}_1} - \tilde{\mathbf{C}_2})(\mathbf{x}) \|_{2}}{\|\mathbf{x}\|_2} \\ &\ge \frac{\sqrt{\text{Area}(\mathcal{P}_1 \ominus \mathcal{P}_2)}(1 - \sqrt{\delta})}{\sqrt{\text{Area}(\mathcal{R})}} - 2\sqrt{\varepsilon}.
\end{align}

\subsection{Proof of Theorem \ref{thm:map_tree}}
For convenience, we restate the theorem:
\begin{theorem*}
Suppose the Laplacians of $S_1, S_2$ have the same eigenvalues, none of which are repeating. Let $\mathcal{M}$ denote the tree constructed from \emph{all isometries} between $S_2 \rightarrow S_1$, and suppose that the refinement algorithm $\mathcal{R}$ is \emph{complete}. Then the subtree of $\mathcal{M}$ until level $\kappa$ will coincide with the tree built from the output of Algorithm A.
\end{theorem*}
\begin{proof}
Under the assumptions of the theorem, any pointwise isometry $T_{12}$ will induce a \emph{diagonal} functional map with values that are $\pm 1$. This is simply because an isometry preserves geodesics (edge lengths in the discrete setting) and thus the local area elements, which implies that the corresponding functional map must be both orthonormal and commute with the Laplace-Beltrami operators. Since none of the eigenvalues are repeating the corresponding functional map must be diagonal.

To prove Theorem \ref{thm:map_tree} suppose there exists a node $n$ in $\mathcal{M}$ which is not present in the tree built from the output of Algorithm A (we denote that tree by $\mathcal{M}_A$). Let $d_n$ represent the diagonal of the functional map corresponding to $n$. Note that $d_n$ is a vector of size $\kappa$. Now for every node $m$ in  $\mathcal{M}_A$ let $d_m$ encode the diagonal of its functional map, and let $p(d_m)$ be the first entry of $d_m$ that does not equal the corresponding entry of $d_n$. In other words $p(d_m) = \min_i \text{s.t. } d_m(i) \ne d_n(i)$. Let $m*$ be the node maximizing $p(d_m)$, i.e. $m* = \argmax_m p(d_m)$, and let $p_{\max} = p(d_{m*})$. Now assume w.l.o.g. that $d_n(p_{\max}) = 1$ whereas $d_{m*}(p_{\max}) = -1$. However, by construction of Algorithm A, to create the node $m*$ it must have considered an initialization consisting of a functional map of size whose diagonal equals to vector $v_1$, of size $p_{\max}$where $v_1(i) = d_{m*}(i), i=1..p_{\max}$, and, again by construction, must have tested the initialization consisting of a functional map whose diagonal equals to $v_2$, where $v_2(i) = d_{m*}(i), i=1..p_{\max}-1$ and $v_2(p_{\max}) = 1$. Note $\mathcal{M}_A$ does not contain nodes $m$, s.t. $d_m(i) = v_2(i), 1..p_{\max}$. However, we supposed that $n$ exists in $\mathcal{M}$, and therefore there must exist at least one isometry whose functional maps has as principal submatrix a diagonal matrix with values equal to $v_2$. But this contradicts the assumption that refinement method $\mathcal{R}$ is \emph{complete}. Therefore, any node in $\mathcal{M}$ must exist in $\mathcal{M}_A$. Conversely, since every node in $\mathcal{M}_A$ corresponds to some isometry it must also exist in $\mathcal{M}$, which proves the theorem.
\end{proof}

%% file: additional_results.tex
\begin{algorithm}[!t]
\DontPrintSemicolon
\SetKwData{Left}{left}\SetKwData{This}{this}\SetKwData{Up}{up}
\SetKwFunction{Union}{Union}\SetKwFunction{FindCompress}{FindCompress}
\SetKwInOut{Input}{Input}\SetKwInOut{Output}{Output}
\Input{A pair of shapes $S_i$ (i = 1,2) with corresponding eigenfunctions $\{\phi_k^{S_i}\}$; initial maps $\Pi_{12}$ and $\Pi_{21}$}
\Output{Refined maps $\Pi_{12}$ and $\Pi_{21}$}
\For{$k = k_{\text{ini}}:k_{\text{step}}:k_{\text{fianl}}$}{
$\Phi_1 = \Phi_k^{S_1},\quad \Phi_2 = \Phi_k^{S_2}$\;
$C_{12} = \Phi_2^{\dagger}\Pi_{21}\Phi_{1},\quad C_{21} = \Phi_1^{\dagger}\Pi_{12}\Phi_{2}$\;
$\Pi_{12} =\text{NNsearch}\big(\Phi_2 C_{21}^T, \Phi_1\big),\quad $
$\Pi_{21} =\text{NNsearch}\big(\Phi_1 C_{12}^T, \Phi_2\big)$\;
$C_{12}= \begin{pmatrix}\Phi_2 \\ \Pi_{12}\Phi_2 \end{pmatrix}^{\dagger} \begin{pmatrix}\Pi_{21}\Phi_1 \\ \Phi_1\end{pmatrix},\quad $
$C_{21}= \begin{pmatrix}\Phi_1 \\ \Pi_{21}\Phi_1 \end{pmatrix}^{\dagger} \begin{pmatrix}\Pi_{12}\Phi_2 \\ \Phi_2\end{pmatrix}$\,
$\Pi_{12} = \text{NNsearch}\big(\begin{pmatrix} \Phi_2C_{21}^T & \Phi_2C_{12}\end{pmatrix}, \begin{pmatrix} \Phi_1 & \Phi_1 \end{pmatrix}\big)$\;
$\Pi_{21} = \text{NNsearch}\big(\begin{pmatrix} \Phi_1C_{12}^T & \Phi_1C_{21}\end{pmatrix}, \begin{pmatrix} \Phi_2 & \Phi_2 \end{pmatrix}\big)$
}
\caption[caption]{Bijective {\zo}}
\label{alg:biject_zoomout}
\end{algorithm}  

\section{Additional results}\label{append:add_res}
\begin{figure}[!t]
    \centering
    \begin{overpic}[trim=8cm 50cm 0cm 28cm,clip,width=1\linewidth,grid=false]{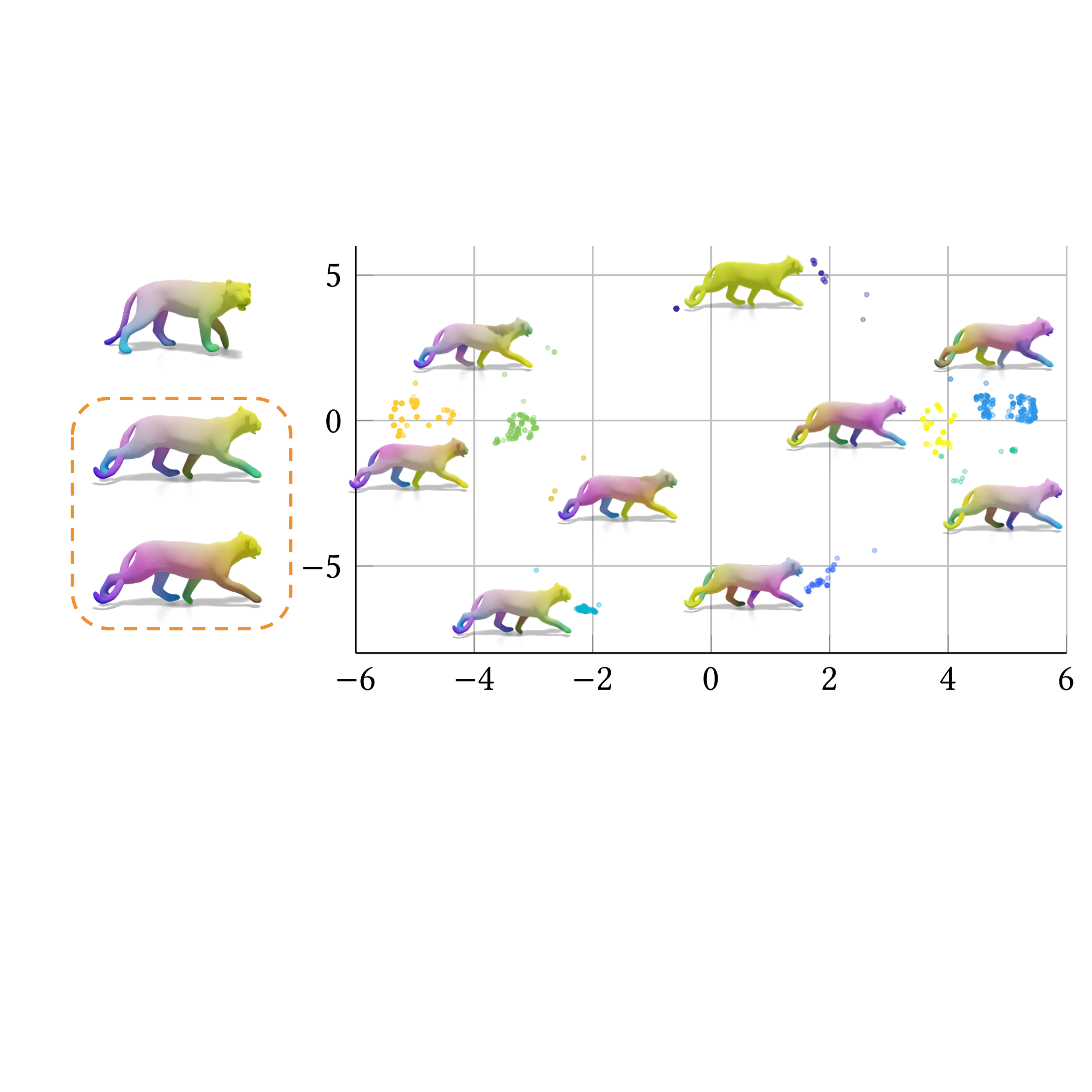}
    \put(8,44){\footnotesize Source}
    \put(8,32){\footnotesize Target}
    \put(-1,5){\footnotesize (Our computed maps)}
    \put(32,46){\footnotesize MDS embedding of 1000 random maps after refinement}
    \end{overpic}\vspace{-12pt}
    \caption{(Left) For this pair of tigers, we apply our method and obtain the direct and the symmetric map visualized via color transfer. (Right) We randomly generate 1000 point-wise maps and apply \zo for refinement. We then embed the refined maps in 2D with MDS. We apply a simple K-means to cluster the 2D points and color each point/map accordingly. For each cluster, we show one representative map that is closest to the direct/symmetric map.}
\label{fig:randIni_map}
\end{figure}

\begin{figure}[!t]
    \centering
    \begin{overpic}[trim=0cm 0cm 0cm 0cm,clip,width=1\linewidth,grid=false]{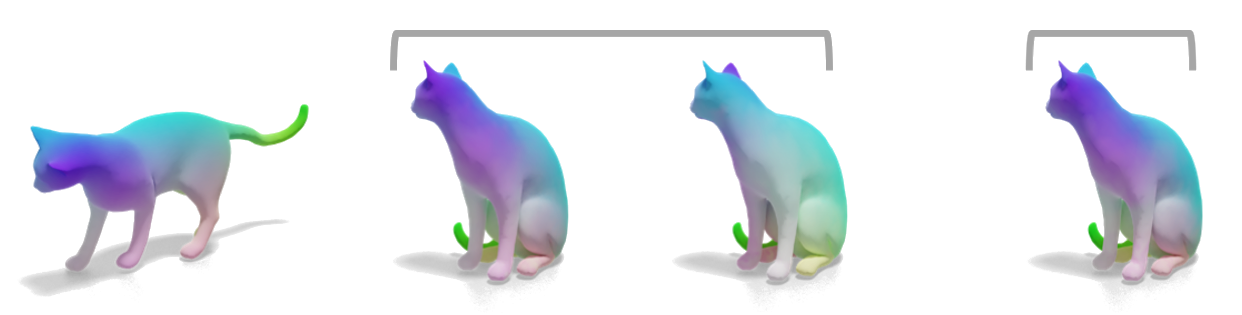}
    \put(8,24){\footnotesize Source}
    \put(38,24){\footnotesize MapTree (\textbf{ours})}
    \put(76,24){\footnotesize randIni + {\zo}}  
    \end{overpic}\vspace{-12pt}
    \caption{For the cat pair, our method obtains both the direct map and the symmetric map simultaneously, while the baseline, applying {\zo} to 500 random initial maps only contains a direct map.}
\label{fig:eg_cat}
\end{figure}

\begin{figure}[!t]
    \centering
    \begin{overpic}[trim=0cm 0cm 0cm 0cm,clip,width=1\linewidth,grid=false]{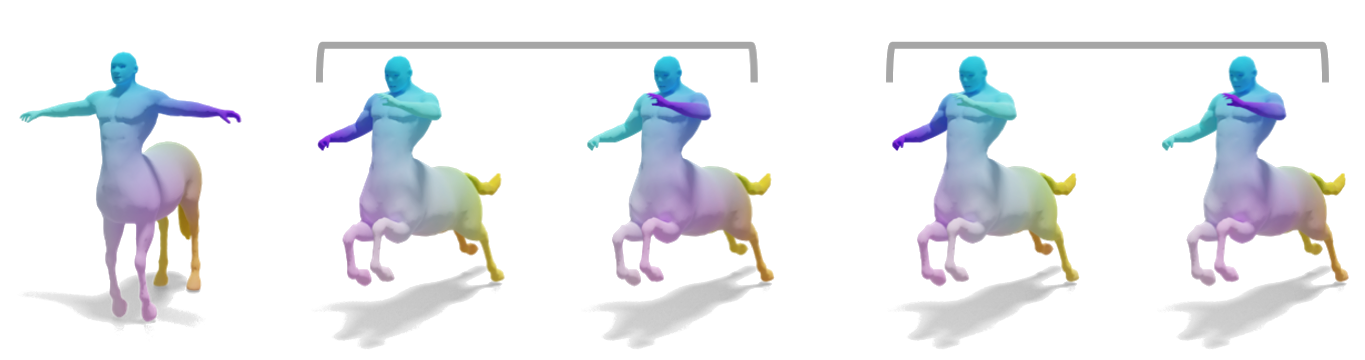}
    \put(5,24){\footnotesize Source}
    \put(30,24){\footnotesize MapTree (\textbf{ours})}
    \put(65,24){\footnotesize randIni + {\zo}}  
    \end{overpic}\vspace{-12pt}
    \caption{For the example of centaurs, the baseline and our method both obtain the direct map and the symmetric map.}
\label{fig:eg_centaur}
\end{figure}

In Algorithm~\ref{alg:biject_zoomout}, we give more details about the bijective {\zo} that is used in our method. Fig.~\ref{fig:randIni_map} shows the corresponding map visualization as discussed in Fig.~\ref{fig:randIni_seg}. Fig.~\ref{fig:eg_cat} and Fig.~\ref{fig:eg_centaur} show two examples of multi-solution shape matching, where we compare our method to a simple baseline: applying {\zo} to 500 random pointwise maps.
In the following, we show more qualitative and quantitative results of our method.

\subsection{Self-symmetry detection}
\begin{figure}[!t]
\centering
\begin{overpic}[trim=11cm 0cm 12cm -3cm,clip,width=0.99\linewidth,grid=false]{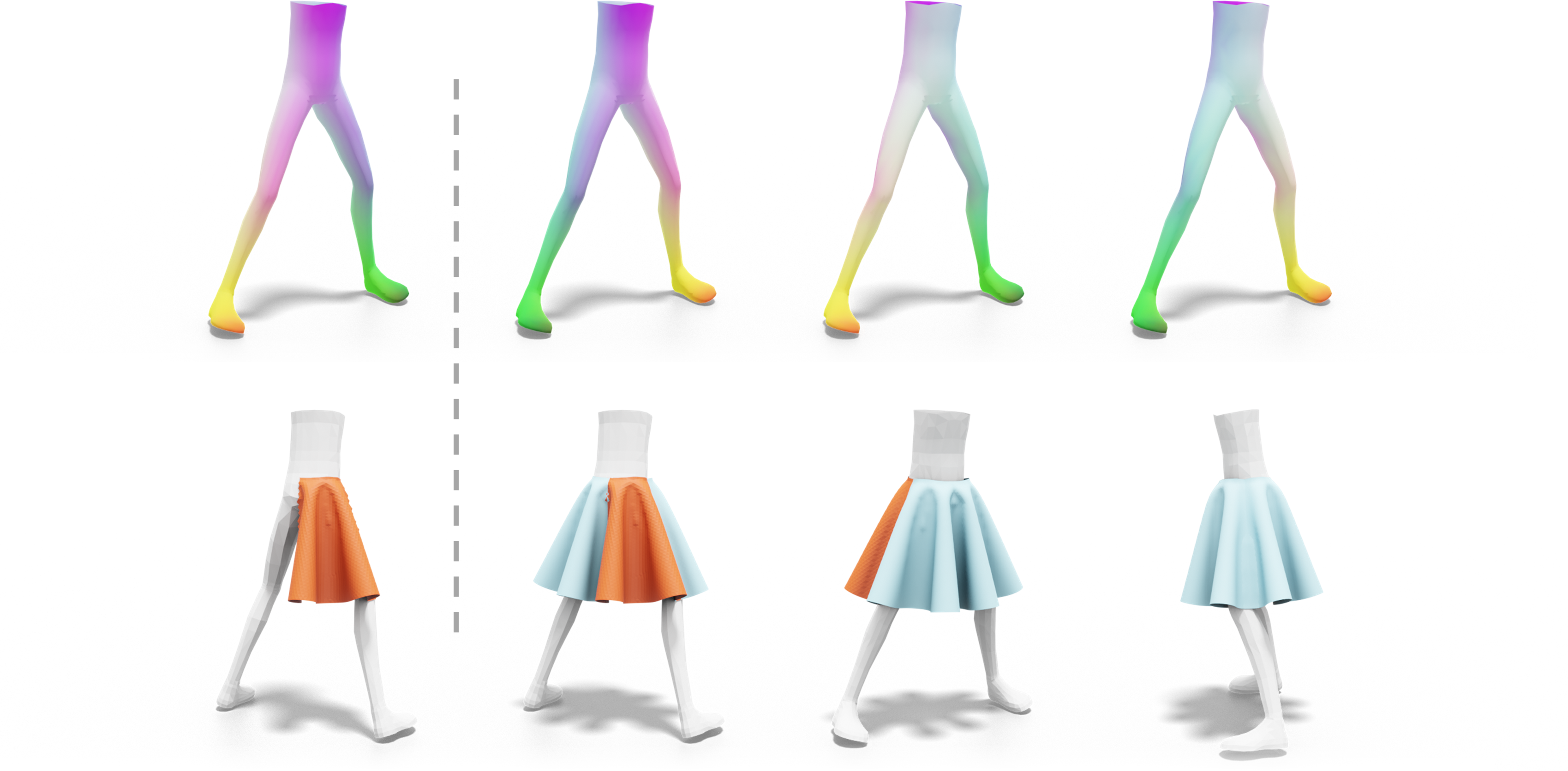}
\put(6.5,70.5){\footnotesize Input}
\put(5,67.5){\footnotesize (Source)}
\put(30,70.5){\footnotesize left-to-right}
\put(31,67.5){\footnotesize symmetry}
\put(55.5,70.5){\footnotesize back-to-front}
\put(57.5,67.5){\footnotesize symmetry}
\put(81.5,70.5){\footnotesize left-to-right \&}
\put(82,67.5){\footnotesize back-to-front}
\put(2, 33){\footnotesize Input garment}
\put(26,33){\footnotesize Completed garment}
\put(58,33){\footnotesize front view}
\put(83,33){\footnotesize side view}
\end{overpic}
\vspace{-10pt}
\caption{\emph{Top}: given an input shape, we compute three different self-symmetric maps visualized via color-transfer.
\emph{Bottom}: given a small piece of garment, we used the computed symmetries to complete the garment. We then show the completed garment in three different views where the input garment is highlighted as red and the automatically generated garment is colored blue. The mesh and garment are from the work~\cite{WangSFM19Garment}.}\label{fig:res:garment}
\vspace{-6pt}
\end{figure}
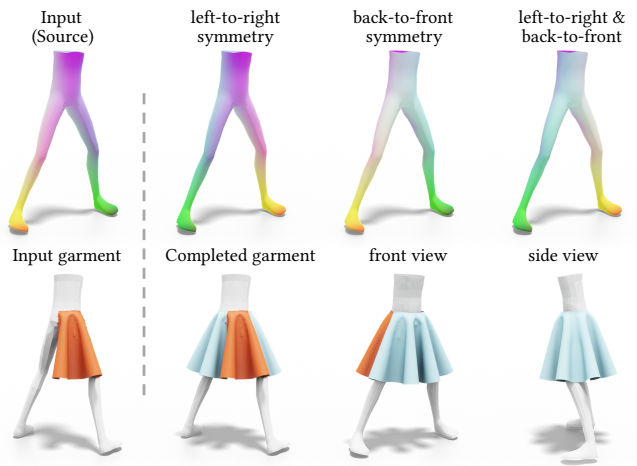

Fig.~\ref{fig:res:garment} shows a use case of how the detected self-symmetries can help the modeling process. To model a symmetric model, in general the artist need to manually setup the symmetry axes during the sculpting. However, it is not trivial to setup such symmetry axes when we want to have the model being symmetric w.r.t. another shape. For example, as shown in Fig.~\ref{fig:res:garment}, we would like to model a dress that is symmetric w.r.t. the given legs shape. In this case, we can use our method to detect the self symmetries of the legs shape and use the obtained symmetries to complete the input garment and make it symmetric from back to front and from left to right w.r.t. the legs shape. 

\setlength{\tabcolsep}{0.8em}
\begin{table}[!t]
\caption{Self-symmetry detection on FAUST dataset with 100 human shapes.}
\label{tb:res:faust:self-symm}\vspace{-6pt}
\footnotesize
\begin{tabular}{|l|c|c|c|}
\hline
\multicolumn{1}{|c|}{Method \textbackslash Measurement} &
  \begin{tabular}[c]{@{}c@{}}Accuracy \\ ($\times 10^{-3}$)\end{tabular} &
  \begin{tabular}[c]{@{}c@{}}Geodesic distortion\\ ($\times 10^{2}$)\end{tabular} &
  \begin{tabular}[c]{@{}c@{}}Runtime \\ (s)\end{tabular} \\ \hline\hline
BIM                           & 65.4          & 2.35                 & 34.6 \\
GroupRep                      & 224           & 10.2                 & 8.48 \\
IntSymm                       & 33.9          & 2.61                 & 1.35 \\
OrientRev (Ini)               & 68.0          & 5.52                 & 0.59 \\
Ini + ICP                     & 38.3          & 3.94                 & 13.3 \\
Ini + BCICP                   & 29.2          & 2.33                 & 195  \\
Ini + ZoomOut                 & 16.1          & 1.86                 & 22.6 \\ \hline\hline
\textbf{Ours} - GT selection & \textbf{15.3} & \textbf{1.81} (2.09) & 61.0 \\ \hline
\end{tabular}
\end{table}

\subsection{Correspondences between a shape pair}
\begin{figure}[!t]
    \centering
    \input{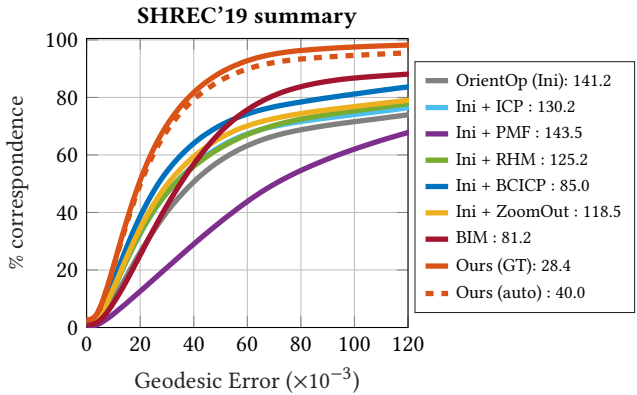}\vspace{-12pt}
    \caption{Shape matching on SHREC'19 Challenge. The summary curves of different methods average over 430 shape pairs.}
    \label{fig:res:shrec19:curve}
\end{figure}

\begin{figure}[!t]
\begin{overpic}[trim=0cm 0cm 0cm -1cm,clip,width=1\linewidth,grid=false]{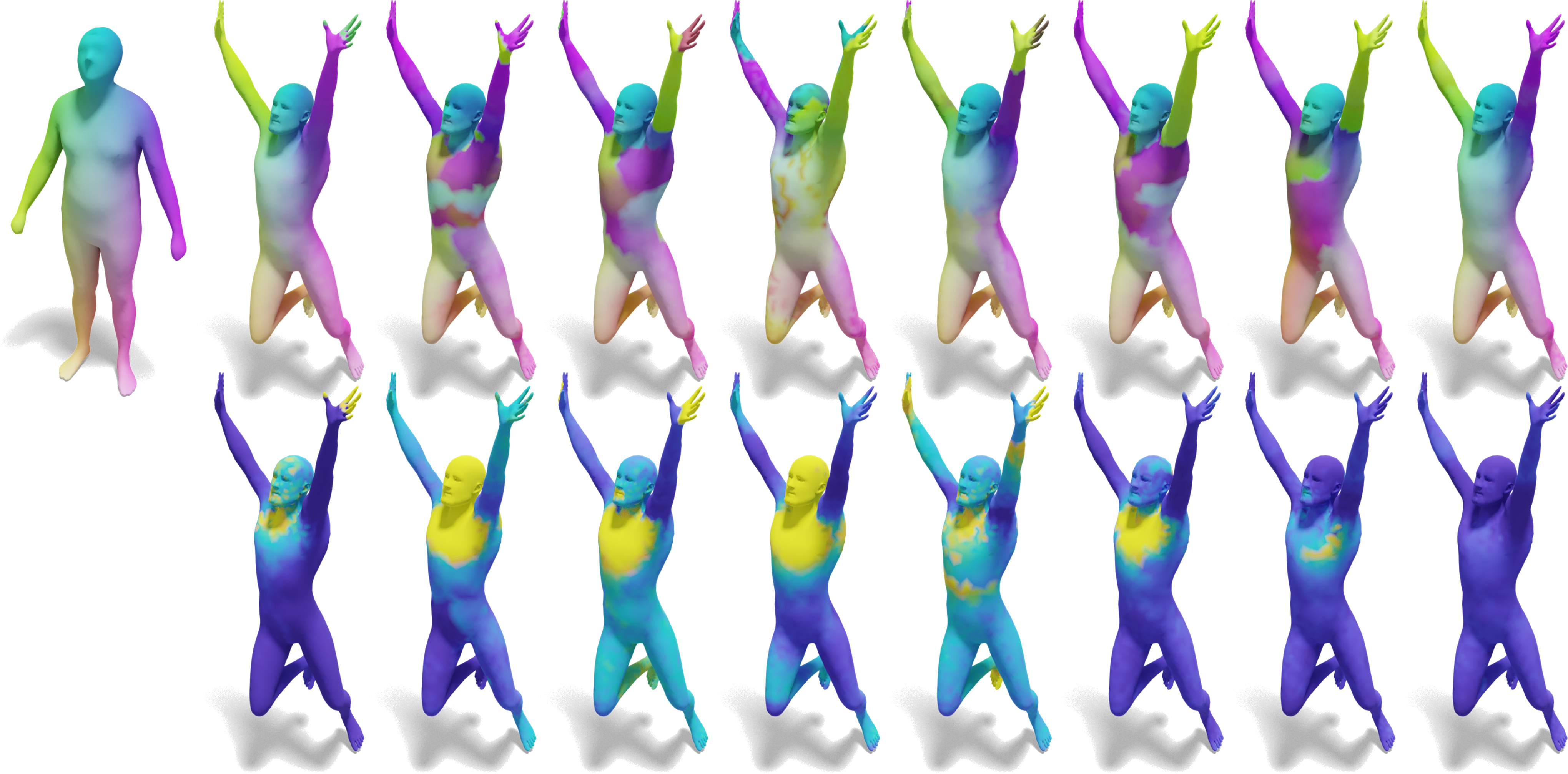}
    \put(2,50){\footnotesize Source}
    \put(15,50){\footnotesize BIM}
    \put(25,50){\footnotesize Orient}
    \put(38,50){\footnotesize ICP}
    \put(48,50){\footnotesize PMF}
    \put(58,50){\footnotesize BCICP}
    \put(69,50){\footnotesize RHM}
    \put(77.5,50){\footnotesize {\zo}}
    \put(92,50){\footnotesize \textbf{Ours}}
    \put(1,15){\footnotesize Geodesic}
    \put(0.6,12){\footnotesize Distortion}
\end{overpic}\vspace{-6pt}
\caption{An example pair from SHREC'19. \emph{Top}: we visualize the computed maps from different methods. \emph{Bottom}: we visualize the geodesic error of each map, where blue indicates a lower error, and yellow indicates a larger error.}
\label{fig:shrec19:geodist}
\end{figure}

\begin{figure}[!t]
    \centering
    \begin{overpic}[trim=0cm 0cm 0cm 0cm,clip,width=0.99\linewidth,grid=false]{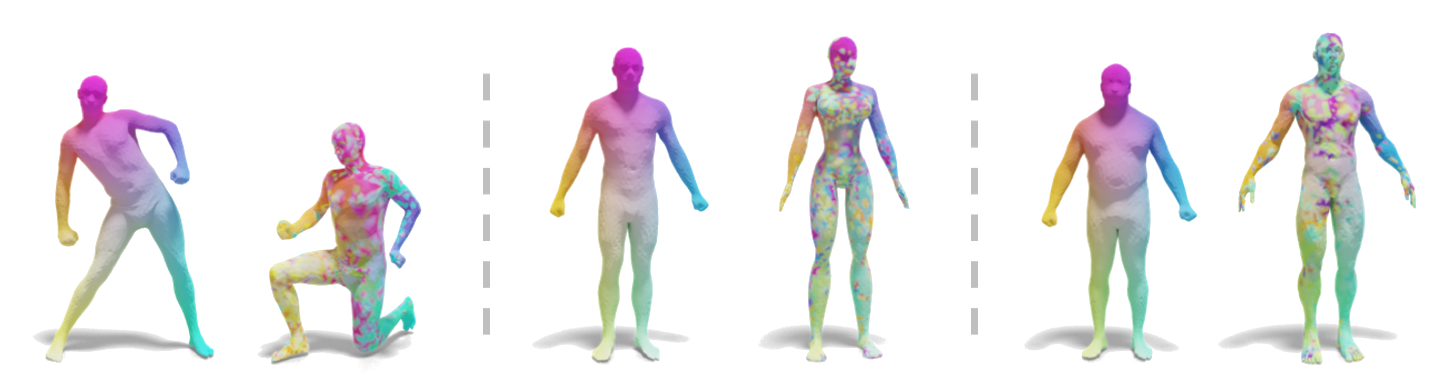}
    \put(9,26){\footnotesize Example 01}
    \put(45,26){\footnotesize Example 02}
    \put(78,26){\footnotesize Example 03}
    \end{overpic}\vspace{-6pt}
    \caption{\revised{Example maps generated by the genetic algorithm~\cite{sahilliouglu2018genetic}. For each example pair, we show the source shape on the left, and visualize the map on the target shape on the right via color transfer.} }
    \label{fig:eg_genetic}
\end{figure}

In Table~\ref{tb:res:shrec19_maptree} we report the accuracy and geodesic distortion of the maps obtained by different methods on 430 shape pairs from the SHREC'19 dataset. Here we show the cumulative curve of the errors in Fig.~\ref{fig:res:shrec19:curve}. Fig.~\ref{fig:shrec19:geodist} shows an example shape pair from the SHREC'19 dataset, for which we visualize both the computed maps and the geodesic distortion. 
\revised{Fig.~\ref{fig:eg_genetic} visualizes three maps generated by the genetic method~\cite{sahilliouglu2018genetic} on shape pairs from the SHREC'19 dataset.}

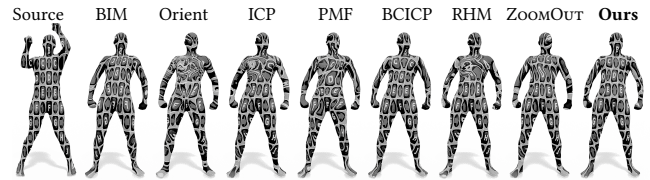
\begin{figure}[!t]
\centering
\begin{overpic}[trim=0cm 0cm 0cm -2cm,clip,width=0.99\linewidth,grid=false]{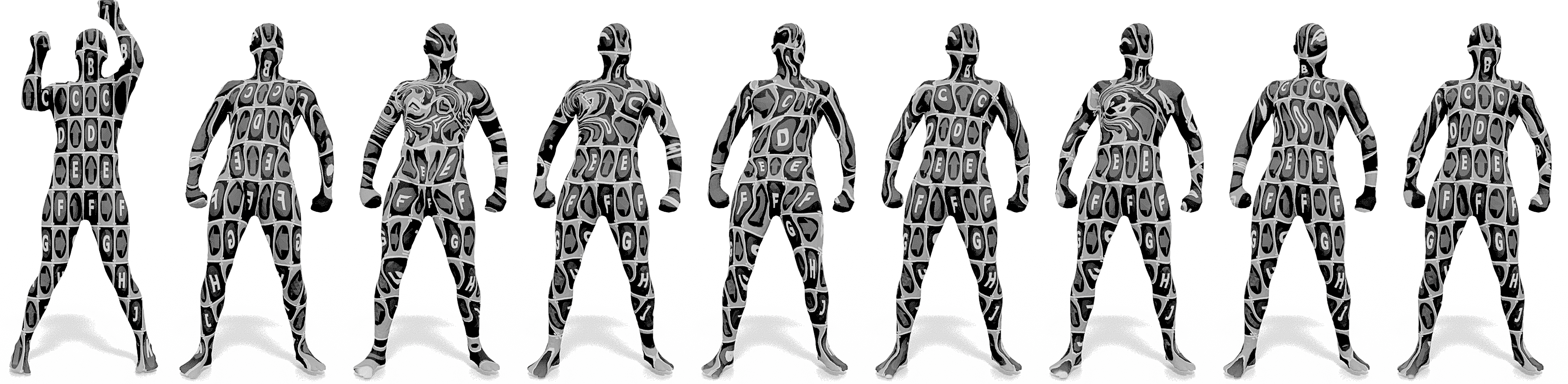}
    \put(0,25){\footnotesize Source}
    \put(13,25){\footnotesize BIM}
    \put(23,25){\footnotesize Orient}
    \put(37,25){\footnotesize ICP}
    \put(48,25){\footnotesize PMF}
    \put(58,25){\footnotesize BCICP}
    \put(69,25){\footnotesize RHM}
    \put(77.5,25){\footnotesize {\zo}}
    \put(92,25){\footnotesize \textbf{Ours}}
\end{overpic}
\vspace{-6pt}
\caption{Here we show an example pair from SCAPE dataset and visualize the obtained maps via texture transfer. }
\label{fig:res:scape_texture}
\end{figure}

\setlength{\tabcolsep}{0.6em}
\begin{table}[!t]
\caption{We test 210 shape pairs from SCAPE dataset and report the average accuracy and runtime for different baseline methods.}
\label{tb:res:scape}\vspace{-3pt}
\footnotesize
\begin{tabular}{|l|c|c|}
\hline
\multicolumn{1}{|c|}{Method / Measurement} & Avg. Error $\big(\times 10^{-3}\big)$ & Avg. Runtime (s)\\ \hline\hline
BIM~\cite{kim2011blended} & 107.6 & 105.5\\
OrientOp (Ini)~\cite{ren2018continuous} & 106.6  & 13.89 \\
Ini + ICP~\cite{ovsjanikov2012functional}  & 71.39 & 13.90\\
Ini + PMF~\cite{vestner2017product} & 119.7 & 385.8 \\
Ini + BCICP~\cite{ren2018continuous} & \textbf{55.92} & 287.1 \\
Ini + RHM~\cite{ezuz2018reversible} & 67.82  &  26.6\\
Ini + {\zo}~\cite{zoomout} & 61.90 &  1.35 \\\hline\hline
\textbf{MapTree (Ours)} - GT selection & \textbf{29.3} & 60.2 \\
\textbf{MapTree (Ours) - Auto selection} & \textbf{42.9} & 60.2 \\ \hline
\end{tabular}
\end{table}

We also perform a similar test to 210 shape pairs from the SCAPE dataset (see Table~\ref{tb:res:scape}). Similarly, we performed a simple ablation study and show that the map tree with our refinement method is more accurate and efficient than using the OrientOp initialization plus the {\zo} refinement. We also use the cycle consistency to automatically select the maps with the preferred direct orientation from our output 5 maps, and the corresponding accuracy is 23\% better than the best baseline. Fig.~\ref{fig:res:scape_texture} shows a qualitative example of the SCAPE shape pair.

\begin{figure*}[!t] 
    \centering
    \vspace{6pt}
    \begin{overpic}[trim=2cm 2cm 3cm 0cm,clip,width=1\linewidth,grid=false]{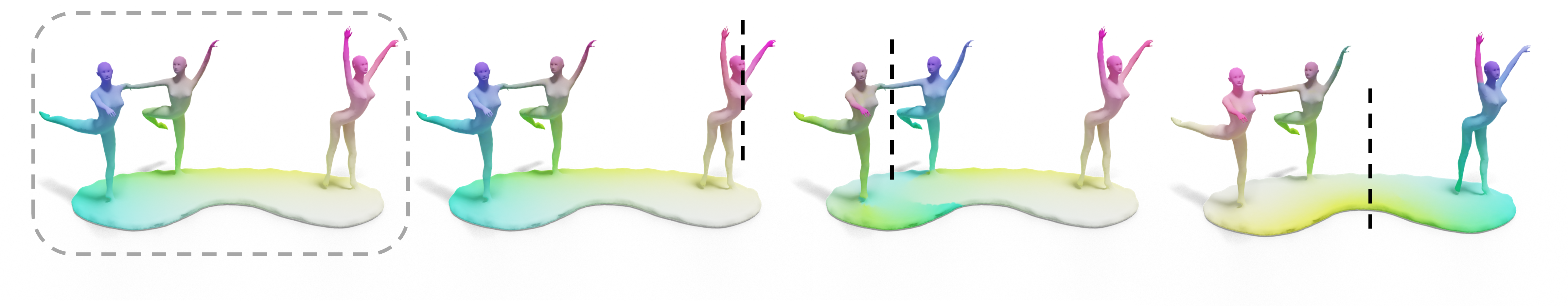}
    \put(1,19){Input: three dancing girls A,B,C}
    \put(2,12){(A)}
    \put(12,12){(B)}
    \put(18,10){(C)}
    
    \put(40,18.5){Symm Axis \circled{1}}
    \put(56.5,17.5){\circled{2}}
    \put(88.5,14.5){\circled{3}}
    \end{overpic}
    \vspace{-1cm}
    \caption{In this example three dancing girls are connected by a base, which makes the overall shape not symmetric. We apply our self-symmetry exploration algorithm directly to obtain self-maps, visualized via color transfer. Note that the partial symmetric axis (1) detected the symmetry of the upper body of the girl (C); the symmetry axis (2) detected the partial symmetry between the girl A and B; the symmetry axis (3) detected a more global partial symmetry, where the girl (A) and (C) are mapped to each other, and (B) is mapped to itself from left-to-right. 
    }
    \label{fig:res:eg_ballet}
    \vspace{12pt}
\end{figure*}

\begin{figure*}[!t] 
    \centering
    \begin{overpic}[trim=0cm 0cm 1cm -2cm,clip,width=1\linewidth,grid=false]{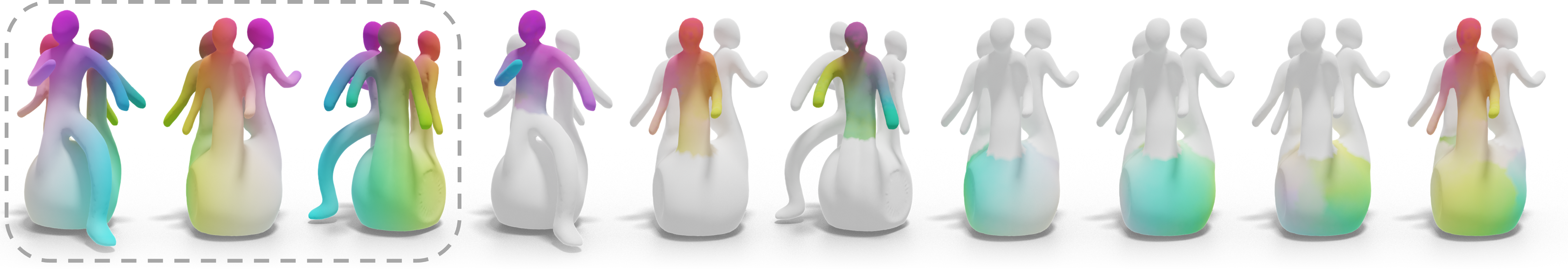}
    \put(4,18){Input (three different views)}
    \put(47,18){A subset of detected partial intrinsic symmetries}
    \end{overpic}\vspace{-6pt}
    \caption{For non-symmetric shapes, some previous methods \cite{xu2009partial,lipman2010symmetry,xu2012multi} can detect the \emph{partial} intrinsic symmetries which are given as \emph{clusters} of points without any exact pointwise matches (please see Fig.14 in~\cite{xu2012multi}). Our method can be applied to non-symmetric shapes as well and obtain the partial symmetric point-wise maps directly.}
    \label{fig:res:eg_memento}
    \vspace{12pt}
\end{figure*}

\begin{figure*}[!t]
    \centering
    \begin{overpic}[trim=0cm 0.8cm 0cm -3.5cm,clip,width=1\linewidth,grid=false]{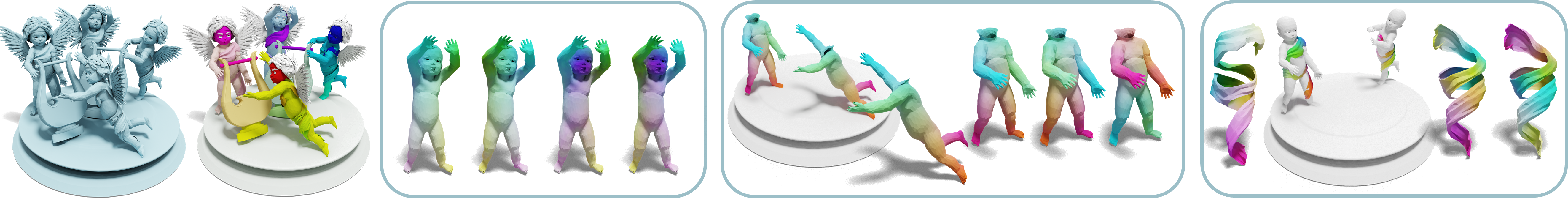}
    \put(3,14){\small Input}
    \put(13,15.5){\small Disconnected}
    \put(13.5,14){\small components}
    \put(24,14.5){\small (A) Self-symm of one component}
    \put(50,14.5){\small (B) maps between two components}
    \put(76.5, 14.5){\small (C) maps between two components}
    \put(26,11.5){\scriptsize (Source)}
    \put(56,10){\scriptsize (Source)}
    \put(78,1.5){\scriptsize (Source)}
    \end{overpic}\vspace{-3pt}
    \caption{Multi-scale symmetry detection. For an input model with multiple objects, we can first decompose it into individual connected components. (A) for a single component, we explore its self-symmetry space. Here we show \emph{three} obtained self-symmetric maps of the baby angel, including the left-to-right, back-to-front, and the double flip self-symmetry. (B) we also explore different maps between two components. In this example, we find the direct map, the symmetric map, and the arm-to-leg correspondences between the body of two angels in different pose. (C) shows another example of finding the correspondences between two pieces of cloth.}
    \label{fig:res:eg_angle}
\end{figure*}
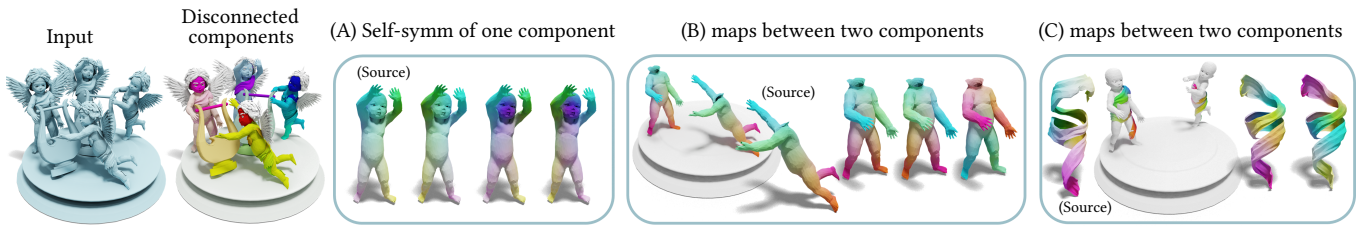

\subsection{Partial symmetry detection}
Fig.~\ref{fig:res:eg_ballet} and Fig.~\ref{fig:res:eg_memento} show other two examples where the input shape is not symmetric intrinsically. In these cases, our method can still be applied directly to obtain some semantic partial symmetries.  
Our method can also be applied to explore the multi-scale structure of an input model with multiple objects. Specifically, we first detect the connected components of the input model, and apply our map exploration method to each component and each pair of components. Fig.~\ref{fig:teaser} shows an example of a shape, in which our method detected different number of symmetry axes across different connected components. Fig.~\ref{fig:res:eg_angle} shows another complex shape, and the results of our method, used to explore how a single connected component can be mapped to itself and to the other components.